\documentclass[10pt, journal, onecolumn, draftclsnofoot]{IEEEtran}
\usepackage{amsmath,amsfonts}
\usepackage{algorithmic}
\usepackage{algorithm}
\usepackage{array}
\usepackage{bm} 
\usepackage{xcolor}
\usepackage[caption=false,font=normalsize,labelfont=sf,textfont=sf]{subfig}
\usepackage{textcomp}
\usepackage{stfloats}
\usepackage{url}
\usepackage{verbatim}
\usepackage{graphicx}
\usepackage{cite}
\usepackage{multirow}
\usepackage{makecell}
\usepackage{microtype}
\usepackage{tabularx}
\usepackage{graphicx}   
\usepackage{etoolbox} 
\usepackage{amssymb} 
\usepackage[T1]{fontenc}    % 核心：使用 T1 编码
\usepackage{balance}	
\def\BibTeX{{\rm B\kern-.05em{\sc i\kern-.025em b}\kern-.08em
		T\kern-.1667em\lower.7ex\hbox{E}\kern-.125emX}}
\usepackage[colorlinks=true,      % 启用彩色链接而非带框链接
			linkcolor=blue,       % 内部链接（如目录、交叉引用）颜色为蓝色
			urlcolor=blue,        % 外部链接颜色为蓝色
			citecolor=blue]{hyperref}

\usepackage{booktabs}        % 学术级横线格式
\usepackage{enumitem}        % 列表格式控制

\newlist{tabitem}{itemize}{1}
\setlist[tabitem]{
	label=\textbullet,
	leftmargin=*, 
	nosep, 
	topsep=1pt, 
	partopsep=0pt, 
	itemsep=2pt,
	before=\footnotesize
}

\begin{document}

\title{Rydberg Atomic Quantum Radio: A Comprehensive Survey From Wireless Communication Perspective}

\author{Yiyue Xiang, Neng Ye,~\IEEEmembership{Senior Member,~IEEE,} Qihao Peng,~\IEEEmembership{Member,~IEEE,}  Junrui Zhao, Qu Luo,~\IEEEmembership{Member,~IEEE,} Kai Yang,~\IEEEmembership{Member,~IEEE,} Jianping An,~\IEEEmembership{Senior Member,~IEEE,} and Pei Xiao,~\IEEEmembership{Senior Member,~IEEE}
        % <-this % stops a space
\thanks{This work was funded by the National Natural Science Foundation of China under grant 62522103. \textit{(Corresponding author: Neng Ye.)}}% <-this % stops a space
\thanks{Yiyue Xiang, Neng Ye, Junrui Zhao and Jianping An are with the School of Cyberspace Science and Technology, Beijing Institute of Technology, Beijing 100081, China (e-mail: yiyuexiang@bit.edu.cn; ianye@bit.edu.cn; junruizhao@bit.edu.cn; an@bit.edu.cn).}
\thanks{Kai Yang is with the School of Information and Electronics, Beijing Institute of Technology, Beijing 100081, China (e-mail: yangkai@bit.edu.cn).}
\thanks{Qihao Peng, Qu Luo and Pei Xiao are with the 5G and 6G Innovation Centre, Institute for Communication Systems (ICS), University of Surrey, GU27XH Guildford, U.K. (e-mail: q.peng@surrey.ac.uk; q.u.luo@surrey.ac.uk; p.xiao@surrey.ac.uk).}
}

%\markboth{Submitted to IEEE Communications Surveys \& Tutorials }%
%{Shell \MakeLowercase{\textit{\textit{et al.}}}: A Sample Article Using IEEEtran.cls for IEEE Journals}


\maketitle

\begin{abstract}
	
Next-generation space-air-ground-sea integrated networks (SAGSIN) impose unprecedented demands on advanced radio frequency (RF) receivers for full-spectrum agility, ultra-high sensitivity, and anti-jamming resilience, pushing conventional electronic receivers to their physical limits. 
To address these challenges, the Rydberg atomic quantum (RAQ) radio has emerged as a  promising quantum-enabled receiver paradigm that directly maps electromagnetic fields onto atomic quantum states, %offering an alternative to conventional RF front ends.
offering an alternative to alleviate bottlenecks of conventional RF front ends.
%offering an alternative beyond conventional RF front ends.
To provide a clear research roadmap, this survey presents a comprehensive review of RAQ radios by bridging atomic physics and wireless communications. 
Specifically, we first introduce the underlying quantum mechanisms, representative architectures, and atomic response models of RAQ radio. On this basis, state-of-the-art techniques for enhancing sensitivity, instantaneous bandwidth, and operating frequency are systematically reviewed, with particular emphasis on the inherent trade-offs among these key metrics. 
To connect quantum response with communication theory, we further analyze equivalent channel modeling frameworks for characterizing  systematic performance limits. 
From the wireless communication perspective, some RAQ-enabled advanced technologies including cognitive, interference-resilient, low-frequency and multiple-input multiple-output (MIMO) communications are reviewed, alongside emerging deployment scenarios such as satellite networks, integrated sensing and communications, and reconfigurable intelligent surface-assisted systems. Finally, we identify open challenges and provide potential future directions of RAQ radio to inspire the further exploration.

\end{abstract}


\section{Introduction}
\IEEEPARstart{T}{he} space-air-ground-sea integrated networks (SAGSIN) represent the definitive paradigm shift in the evolution of next-generation wireless systems \cite{guo2022sagsinsecurity,xiao2024sagin6g}. It is envisioned to establish ubiquitous, seamless connectivity across all physical domains, integrating the space tier of geosynchronous, medium, and low earth orbit (GEO/MEO/LEO) satellites, the air layer of unmanned aerial vehicles (UAVs), the terrestrial network of ground terminals, and the maritime segment reaching down to deep-sea submersibles \cite{guo2022sagsinsecurity,xiao2024sagin6g,luo2024leovleo,xiao2022mmwaveuav,lin2023underwater}. This open architecture inherently encounters security threats, such as reconnaissance, interception, and electronic jamming \cite{guo2022sagsinsecurity}, while simultaneously operating across a wide spectrum that spans over 12 orders of magnitude, stretched from Hertz-level extremely low frequencies (ELF) for underwater communication \cite{rowe1974elf} to millimeter-wave Q/V bands for high-capacity satellite links. 
These heterogeneous operational environments impose stringent imperatives on the underlying hardwares, including full-spectrum agility, ultra-weak signal detection, anti-jamming resilience, and compact integration, which have pushed conventional radio frequency (RF) receivers to their physical limits \cite{rossi2016alphasat}.
Incremental iterations in circuit design and fabrication processes are challenging to break through the fundamental constraints of classical physics, rendering conventional RF receivers difficult to satisfy these multi-dimensional requirements.

Reflecting on the century-long evolution of wireless communications, the fundamental operating paradigm of receivers has remained generally consistent across three distinct technological epochs. From the early stage marked by Hertz's seminal electromagnetic resonant-loop prototype in 1887 \cite{hertz1889strahlen}, through the analog receiver era initiated by Marconi's trans-Channel radio broadcasts experiment \cite{bondyopadhyay1995marconi}, and ultimately to the contemporary digital software-defined radios \cite{mitola1995software}, the core physical mechanism of wireless receivers is hardly changed. Based on the forced oscillation of free electrons driven by incident electric fields, conventional RF receiver rely inherently on macroscopic metallic front-end antennas to transduce free-space electromagnetic radiation into alternating conduction currents \cite{griffiths2017introduction,balanis2005antenna}. Once captured, the backend semiconductor components process these signals within the electronic domain, through a cascade of operations including amplification, down-conversion and filtering\cite{moghaddasi2020multifunction,razavi2011rf}.

This classical architecture of field-to-current conversion and electronic signal processing is bounded by several inherent classical physical limits:
\begin{itemize}
    \item \textit{Chu’s Limit on Antenna Size}: The Chu–Harrington limit indicates that electrically small antennas face a fundamental trade-off among antenna size, bandwidth, and radiation efficiency \cite{chu1948physical}. Therefore, efficient and broadband reception at low-frequencies such as 3 MHz remains highly challenging, as its wavelength is on the order of 100 m.

    \item \textit{Thermal Noise Limit on Sensitivity}: 
    In conventional semiconductor devices, the irregular thermal motion of free electrons at room temperature introduces an inherent -174 dBm/Hz noise floor \cite{cui2024rydberg}. Component and transmission-line losses further set the practical sensitivity floor of commercial spectrum-analyzer receivers at approximately the $-150$ dBm level \cite{nist2022atomictelevision}.
    % Since conventional receivers manipulate electrical currents within In the semiconductor devices, the irregular thermal motion of free electrons at room temperature introduces an inherent noise floor of -174 dBm/Hz. In practical engineering, compounding component and transmission line losses typically cause the sensitivity ceiling of commercial receivers to plateau around -120 dBm.
    %This imposes a rigid threshold on the ultimate sensitivity achievable for long-distance, ultra-weak signal detection.

    \item \textit{Hardware Vulnerability}:
    Active semiconductor RF front-ends have severely limited linear dynamic range. Signal inputs exceeding 10 dBm induce severe saturation distortion, while electromagnetic interference (EMI) above 30 dBm can cause irreversible burnout \cite{yang2023sub1ms}. %The linear dynamic range of active semiconductor RF front-ends is severely restricted. Signal inputs exceeding 10 dBm typically induce severe saturation distortion, while electromagnetic interference (EMI) above 30 dBm can cause irreversible physical burnout. This vulnerability limits the anti-jamming resilience of conventional receivers, preventing stable operation in highly contested electromagnetic environments.
\end{itemize}
As these physical boundaries constrain the
development of SAGSIN, there is an urgent need to move beyond the conventional electronics framework and explore new solutions for next-generation wireless reception technologies.

\begin{figure*}[!t]
    \centering
    \includegraphics[width=0.92\textwidth]{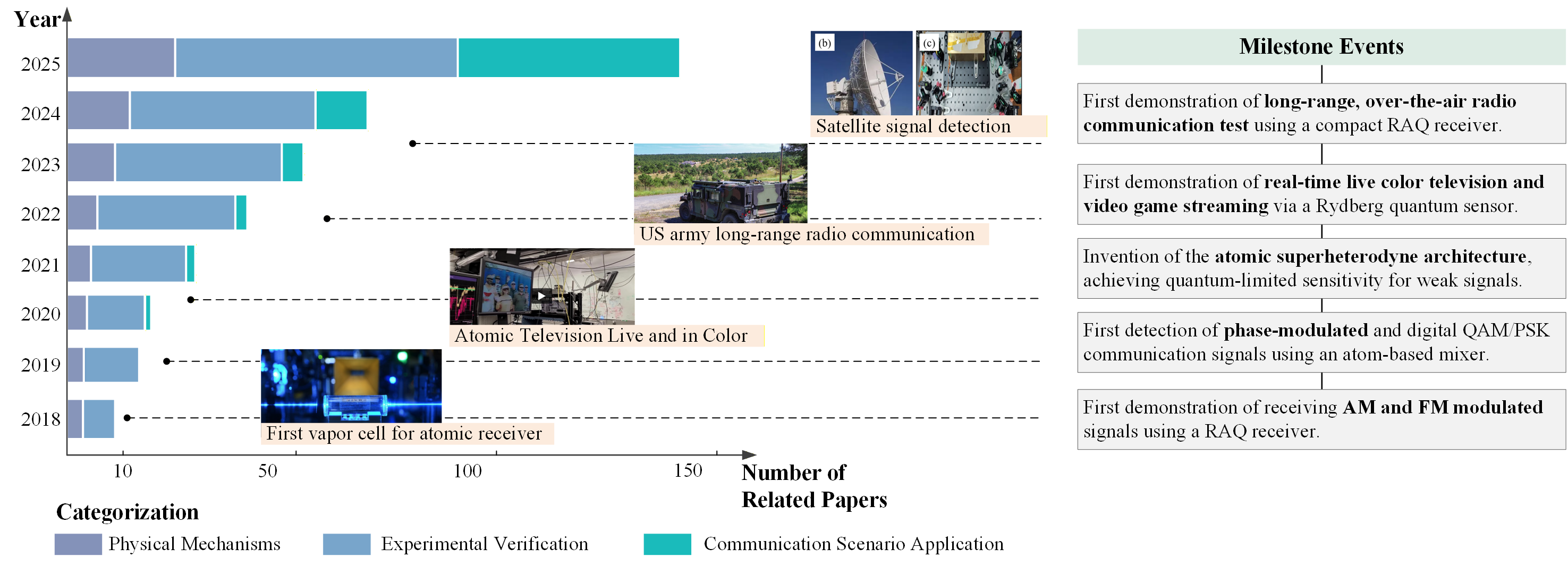}
    \caption{Publication trends and technological milestones of RAQ receivers for wireless communications. First vapor cell for atomic receiver \cite{meyer2018digital}. Atomic Television live and in color \cite{NIST_TV}. US army long-range radio communication \cite{USdefense}. Satellite signal detection \cite{lei2025satellite}.}
    \label{fig_milestone}
\end{figure*}

Motivated by this need, researchers have turned to fundamental physics to seek solutions. Among these efforts, Rydberg atomic quantum (RAQ) radio has emerged as a promising solution and attracted growing attention.
Summarized in Fig. \ref{fig_milestone}, the exploration of RAQ radio for wireless communications accelerated around 2018, when early studies redesigned the RF receivers by replacing metallic antennas with centimeter-scale atomic vapor cells \cite{holloway2014broadband,cox2018quantum,song2018quantum,Song:19}. By using lasers to probe the optical response of atoms to RF fields, these demonstrations verified the feasibility of receiving amplitude-modulated (AM) and frequency-modulated (FM) analog signals \cite{anderson2021atomic,holloway2021multiple,meyer2018digital}.
Following this, RAQ radio rapidly evolved toward more sophisticated communication functions. Atomic mixers enabled the extraction of signal phase information, expanding the supported formats to digital phase-shift keying (PSK) and quadrature amplitude modulation (QAM) \cite{holloway2019detecting, nowosielski2024warm}. To improve weak-signal detection, atomic superheterodyne architectures were developed, pushing sensitivity toward quantum projection noise limits \cite{jing2020atomic}. Furthermore, by manipulating atomic transit times to overcome bandwidth constraints, researchers enabled high-speed data transmission, facilitating applications such as real-time color television streaming \cite{NIST_TV,prajapati2022tv}. 
%As evidenced by the continuous technological iterations and the rapid growth of related publications (Fig. \ref{fig_milestone}), the research focus is gradually transiting from laboratory physics investigations to practical communication engineering. Recent advances have miniaturized the receiver into compact configurations, enabling long-range, over-the-air wireless communications in complex outdoor environments \cite{USdefense,lei2025satellite}.
As evidenced by continuous technological iterations and the rapid growth of related publications (Fig. \ref{fig_milestone}), these advances indicate a gradual shift from laboratory physics demonstrations to communication engineering. Recent compact receiver prototypes have further enabled long-range over-the-air wireless communications in more realistic outdoor environments \cite{USdefense,lei2025satellite}.

Following recent advances, RAQ radio is defined in this survey as a quantum-enbaled radio paradigm that exploits the coherent coupling between free-space electromagnetic fields and resolvable transitions of discrete atomic quantum states for wireless communications. Among the existing studies, RAQ receivers for signal reception are the dominant realization of this paradigm. Their core innovation lies in the reconstruction of the reception mechanism, namely, instead of converting electromagnetic fields into conduction currents, RAQ receiver directly maps free-space electromagnetic information onto the quantum-state evolution of an atomic ensemble, from which the signal can be recovered through optical readout. This field-to-state mapping alleviates several classical electronic bottlenecks and endows RAQ receivers with distinctive advantages:

\begin{itemize}
    \item \textit{Ultra-Wideband Frequency Coverage}: 
    %A centimeter-scale atomic vapor cell can response to signals with frequencies spanning from Hz to THz regime \cite{jau2020vapor,meyer2020assessment}, thereby breaking through the constraints imposed by Chu’s limit.
    A centimeter-scale atomic vapor cell can detect signals across a spectrum spanning from Hz to Terahertz (THz) regime \cite{jau2020vapor,meyer2020assessment}, thereby bypassing the constraints imposed by Chu’s limit.

    \item \textit{High Sensitivity Beyond the Thermal Noise Limit}: %Theoretically, atomic radio can bypasses the room-temperature thermal noise floor that plagues conventional electronics, achieving quantum-limited sensitivity \cite{cox2018quantum}.
   The theoretical sensitivity of atomic radio is bounded by the quantum limit, with its noise floor falling below the room-temperature thermal noise of electronics receiver \cite{cox2018quantum}.

    \item \textit{Inherent Anti-Interference Resilience}: 
    %The stringent frequency selectivity of atomic energy-level transitions mechanism provides innate anti-interference resilience \cite{fancher2021rydberg}, guaranteeing robust and stable operation even in highly contested electromagnetic environments.
    The stringent frequency selectivity in atomic energy-level transitions introduces a unique filtering effect \cite{fancher2021rydberg} that enables inherently rejection of out-of-band signals, providing potential for sustaining reliable communications even within highly congested electromagnetic scenarios.

    \item \textit{Miniaturized and Unified Front-End Architecture}: The direct atomic-optical readout mitigates the need for separate band-specific RF front-ends, thus facilitating the feasibility of simultaneous and cross-band signal reception.
\end{itemize}
To sum up, these capabilities position RAQ receivers as an compelling  candidate to meet the stringent demands of the next-generation SAGSIN.

\subsection{Related Works}
\begin{table*}
	\centering
	\caption{Comparison of Related Works on RAQ radio}
	\label{tab:survey_comparison}
	\small
	\setlength{\tabcolsep}{3pt}
	\renewcommand{\arraystretch}{1.3}
	\resizebox{\textwidth}{!}{
	\begin{tabular}{|c|*{8}{c|}|c|}
		\hline
		\textbf{Ref} & 
		\rotatebox{90}{\shortstack{Atomic Physics\\Fundamentals}} &
		\rotatebox{90}{\shortstack{System\\ Architecture}} &
		\rotatebox{90}{\shortstack{Atomic Response \\ Model}} &
		\rotatebox{90}{\shortstack{Metric Enhancement~~\\\& Trade-off}} &
		\rotatebox{90}{\shortstack{Equivalent\\Channel Model}} &
		\rotatebox{90}{\shortstack{Impairments\\and Interference}} &
		\rotatebox{90}{\shortstack{Rydberg-Enhanced\\Communication \\ \& Sensing}} &
		\rotatebox{90}{\shortstack{~~~Application\\~~Scenarios}} &
		\textbf{Core Contribution} \\
		\hline
		\makecell[c]{\cite{artusio2022modern}} & $\ast$ & $\checkmark$ & $\times$ & $\ast$ & $\times$ & $\times$ & $\ast$ & $\ast$ & \makecell[c]{Hot atoms for RF-field measurements} \\
		\hline
		\makecell[c]{\cite{liu2023electric}} & $\checkmark$ & $\checkmark$ & $\ast$ & $\ast$ & $\times$ & $\times$ & $\ast$ & $\ast$ & \makecell[c]{Electric-field measurements for various frequency bands} \\
		\hline
		\makecell[c]{\cite{schlossberger2024rydberg}} & $\ast$ & $\checkmark$ & $\times$ & $\checkmark$ & $\times$ & $\ast$ & $\ast$ & $\checkmark$ & \makecell[c]{Rydberg atom-based precise electrometry and application} \\
		\hline
		\makecell[c]{\cite{zhang2024rydberg}} & $\ast$ & $\checkmark$ & $\times$ & $\ast$ & $\times$ & $\times$ & $\ast$ & $\ast$ & \makecell[c]{Performance comparison of different RAQ sensing architectures} \\
		\hline
		\makecell[c]{\cite{babusenan2026rydberg}} & $\checkmark$ & $\checkmark$ & $\ast$ & $\checkmark$ & $\times$ & $\checkmark$ & $\ast$ & $\checkmark$ & \makecell[c]{\makecell[c]{Systematic review of Rydberg atom-based RF sensing}} \\		
		\hline
		\makecell[c]{\cite{fancher2021rydberg}} & $\ast$ & $\checkmark$ & $\times$ & $\checkmark$ & $\times$ & $\ast$ & $\ast$ & $\ast$ & \makecell[c]{Performance comparison with traditional receivers} \\
		\hline
		\makecell[c]{\cite{cui2024rydberg}} & $\ast$ & $\ast$ & $\times$ & $\ast$ & $\times$ & $\times$ & $\checkmark$ & $\checkmark$ & \makecell[c]{Numerical simulation of performance metrics} \\
		\hline
		\makecell[c]{\cite{gong2025rydberg}} & $\ast$ & $\checkmark$ & $\times$ & $\ast$ & $\times$ & $\times$ & $\checkmark$ & $\checkmark$ & \makecell[c]{RAQ-SISO and RAQ-MIMO architectures} \\
		\hline
		\makecell[c]{\cite{chen2025new}} & $\ast$ & $\ast$ & $\times$ & $\ast$ & $\ast$ & $\times$ & $\checkmark$ & $\checkmark$ & \makecell[c]{The hybridization between RAQ receiver and ISAC} \\			
		\hline
		\makecell[c]{Our Work} & $\checkmark$ & $\checkmark$ & $\checkmark$ & $\checkmark$ & $\checkmark$ & $\checkmark$ & $\checkmark$ & $\checkmark$ & \makecell[c]{Systematic review of RAQ receivers in wireless communication} \\
		\hline
	\end{tabular}
}
	\medskip
	\centering
	$\checkmark$ Primary focus \quad $\ast$ Partial focus \quad $\times$ Not a focus.
\end{table*}

Several review-oriented works on RAQ receivers have been published, mainly covering fundamental principles, receiver architectures, performance comparisons with conventional radio receivers, and representative application scenarios~\cite{fancher2021rydberg,artusio2022modern,liu2023electric,cui2024rydberg,schlossberger2024rydberg,zhang2024rydberg,gong2025rydberg,chen2025new,babusenan2026rydberg}, as summarized in Table~\ref{tab:survey_comparison}.

Regarding the field sensing, Artusio-Glimpse \textit{et al.}~\cite{artusio2022modern} focus on hot-atom Rydberg RF sensors, discussing quantum-interference mechanisms, receiver architectures, performance-enhancement techniques, and applications in RF power measurement, voltage standards, and spectrum analysis.  Liu \textit{et al.}~\cite{liu2023electric} review electric-field measurement methods from direct current (DC) to THz, covering DC sensing, low-frequency vapor-cell effects, Floquet-based intermediate-frequency detection, and high-frequency AT splitting. Schlossberger \textit{et al.}~\cite{schlossberger2024rydberg} provide a survey of Rydberg electrometry with emphasis on sensitivity-enhancement techniques, environmental effects, measurement uncertainty, continuous-frequency detection, and emerging applications such as sensor arrays and portable probes. Zhang \textit{et al.}~\cite{zhang2024rydberg} summarize RAQ receiver architectures, performance characteristics, and advanced methods such as microwave frequency combs, non-equilibrium sensing, and deep-learning-assisted measurements. Babusenan \textit{et al.}~\cite{babusenan2026rydberg} provide a systematic review covering SI-traceable Rydberg electrometry, excitation schemes, sensing and receiver architectures, system-level readout strategies, and performance metrics.

In terms of communication, Fancher \textit{et al.}~\cite{fancher2021rydberg} review atomic electric-field sensors for communications and sensing, introducing their operating principles, receiver architectures, performance capabilities, and applications such as spectrum situational awareness, distributed sensing, and antenna characterization. Cui \textit{et al.}~\cite{cui2024rydberg} compare RAQ receivers with conventional antennas in terms of size, sensitivity, and bandwidth, and further summarize RAQ-enabled communication and sensing techniques, including multi-band sensing, atomic multiple-input multiple-output (MIMO), and quantum-enhanced noise mitigation. Gong \textit{et al.}~\cite{gong2025rydberg} introduce the fundamentals of Rydberg atoms and EIT-based detection, classify receiver architectures, and discuss RAQ-SISO and RAQ-MIMO schemes for integration with classical wireless systems. Chen \textit{et al.}~\cite{chen2025new} focus on sensing and communication applications and propose an ISAC framework using FM and acousto-optic frequency shifters to overcome instantaneous-bandwidth limitations.

\begin{figure*}[!t]
	\centering
	\includegraphics[width=0.95\textwidth]{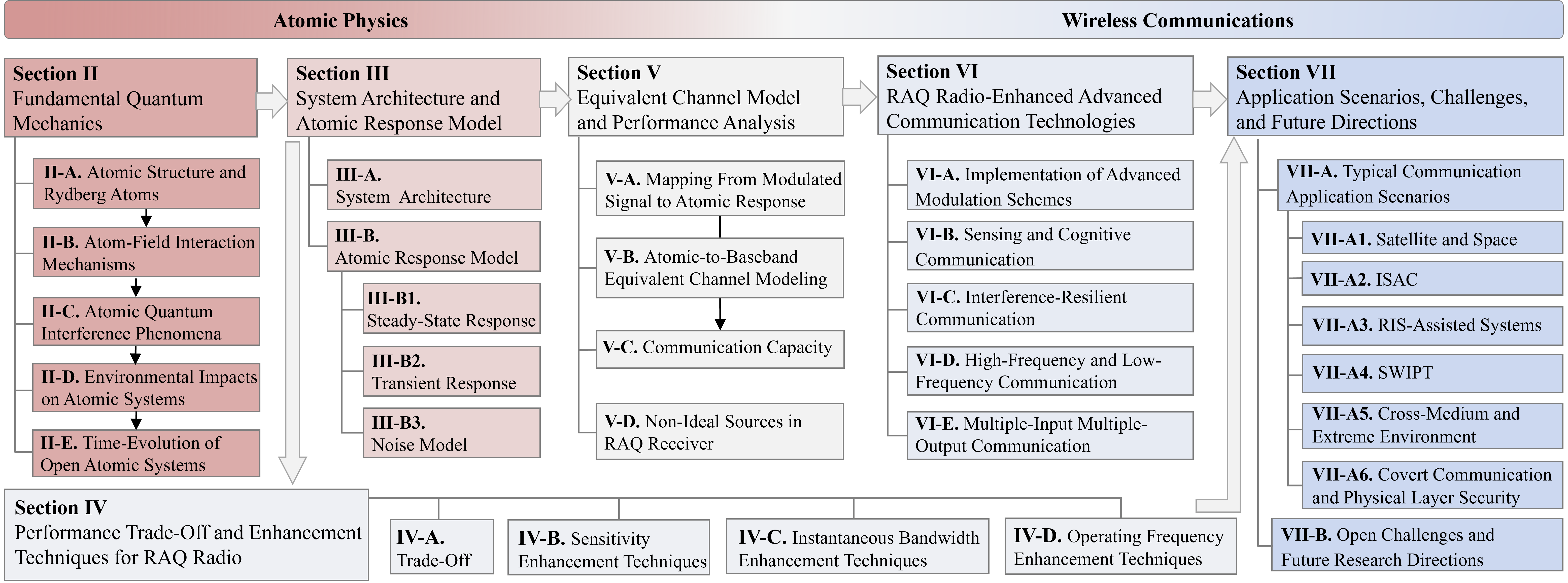}
	\caption{Organization of this paper.}
	\label{fig:organization}
\end{figure*}

%As summarized in Table~\ref{tab:survey_comparison}, some works emphasis atomic physics principles, while others concentrate on practical wireless communication applications. 
As summarized in Table~\ref{tab:survey_comparison}, some works focus on Rydberg-atom-based field measurement, whereas others discuss selected communication aspects.
Despite these efforts, a systematical discussion on the theoretical research bridging atomic quantum physics and wireless communication remains absent, particularly the critical relationship between the physical atomic response and the equivalent communication channel.
%In response, this survey provides a comprehensive and systematic review of RAQ receivers, with particular emphasis on equivalent channel modeling, channel impairments and interference, Rydberg-enhanced communication and sensing techniques, and practical application scenarios.

\subsection{Contributions and Organization}
This paper presents a comprehensive survey of RAQ radios by bridging atomic physics and wireless communications.
Building on the basic principles of atomic quantum mechanics, we systematically review the RAQ radio from its underlying physical mechanisms to communication applications. We not only elaborate on the unique enabling value of RAQ radio for wireless communications, but also analyze how practical engineering imperatives drive the evolution of RAQ radio. This survey aims to provide a clear research roadmap and reference framework for interdisciplinary integration of RAQ radio. 

The main contributions are summarized as follows:
\begin{itemize}
	\item We systematically introduce the principle of RAQ radio, covering the underlying physical fundamentals, system architectures, and operating mechanisms. Notably, we provide the first comparisons of RAQ receivers under three typical quantum interference pathways--$\Lambda$-, $\Xi$-, and V-type configurations, and discuss their scope of applications.
	%--in terms of physical characteristics, performance limits, and application scenarios.
	%, which fills the research gap of emerging atomic structures and practical communication applications.
	
	\item We analyze the intrinsic connection between atomic quantum responses and equivalent communication channels.
	By clarifying the information mapping from incident electromagnetic fields to baseband outputs, and synthesizing modeling approaches based on the optical Bloch equations, we summarize a systematic theoretical framework for the quantitative performance evaluation of RAQ radios.
	
	\item We conduct the comprehensive review of state-of-the-art enhancement techniques and analyze their inherent trade-offs regarding key performance metrics of RAQ receivers, including sensitivity, instantaneous bandwidth and operating frequency, thereby offering essential guidelines for system-level design and implementation.
	
	\item 
	We examine the enabling potential of RAQ radio for future wireless communication systems.
	Specifically, we summarize how RAQ receivers enable emerging communication scenarios, such as satellite networks, integrated sensing and communications, and multi-input multiple-input multiple-output communications. We further identify communication-driven requirements on system integration, intelligent processing, and operational stability, thereby outlining key research directions for future RAQ radio.

\end{itemize}

As shown in Fig.~\ref{fig:organization}, this paper is organized as follows. Section II provides the quantum mechanical foundations necessary for understanding RAQ radios. Section III presents the system architecture and clarifies the physical response models. Section IV reviews recent advances in performance enhancement across three key metrics and their trade-off. Section V bridges quantum physics and communication theory by clarifying equivalent channel models and analyzing performance limits. Section VI explores advanced communication technologies enabled by RAQ radios. Section VII discusses applications scenarios, open challenges, and future directions. 

%\emph{Notations:} 

\section{Fundamental Quantum Mechanics of Rydberg Atoms}
The core principle of RAQ radio lies in the microscopic quantum-state evolution of atomic ensembles, which relies on the interactions between external electromagnetic fields and internal atomic energy levels. In this section, we first introduce the structural properties of Rydberg atoms and the fundamental mechanics of atom-field interactions. Then, we delve into the two core quantum interference phenomena that enable the optical readout of RF signals. Finally, we analyze the environmental impacts on ideal physical models and  introduce the density matrix to model the time-evolution of the open atomic systems.

\subsection{Atomic Structure and Rydberg Atoms}\label{sec:basic-concepts}
In quantum mechanics, the energy of atoms is quantized, which is defined by a series of discrete and stationary energy states occupied by their electrons. These states are characterized by a set of quantum numbers, including the principal quantum number $n$, the orbital angular momentum quantum number $l$, the magnetic quantum numbers $m$, and the electron spin quantum number $s$, which jointly determine the energy of atoms and govern their interaction characteristics with external electromagnetic fields.
Specifically, the principal quantum number $n, \left(n=1,2,3, \cdots \right)$ serves as the primary criterion for defining the mean radial distance from the nucleus, dominating the energy of atoms. The orbital angular momentum quantum number $l, \left(l=0,1,2, \cdots, n-1\right)$ describes the orbital shape of electrons and the magnitude of angular momentum. The projection of this angular momentum along a specific quantization axis, e.g., the direction of magnetic field, is given by the magnetic quantum number $m, \left(m=0, \pm1, \cdots, \pm l \right)$. Finally, the intrinsic spin angular momentum of electrons is described by the spin quantum number $s$, which is fixed at $\pm1/2$.

\begin{figure*}[!t]
	\centering
	\includegraphics[width=\textwidth]{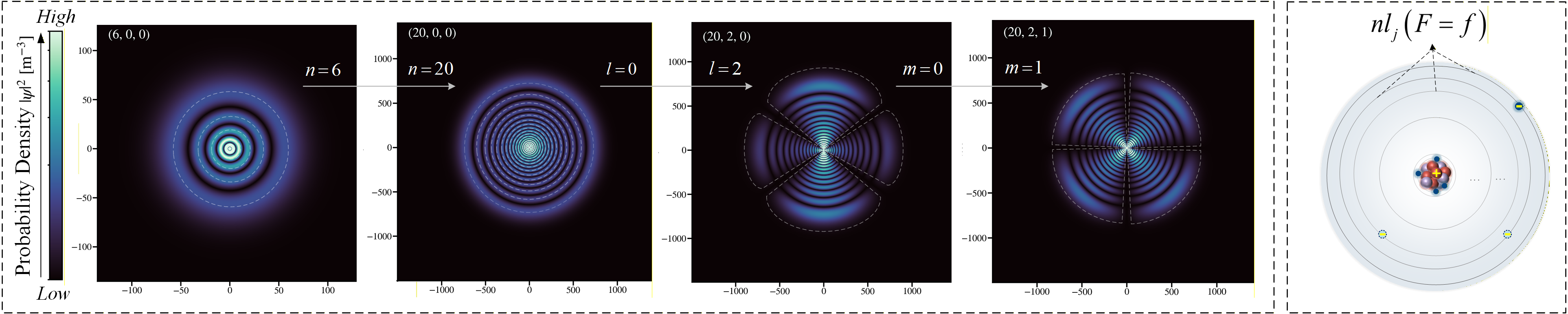} 
	\caption{The electron cloud model of atoms: probability density distributions of the electronic wavefunctions $|\psi_{nlm}|^2$ for varying quantum numbers $(n, l, m)$ (from left to right: $(n,l,m)=(6, 0, 0)$, $(20, 0, 0)$, $(20, 2, 0)$, and $(20, 2, 1)$), alongside a simplified atomic model with quantum state and energy level correspondence.}
	\label{fig:atom}
\end{figure*}

Accordingly, we can represent the energy level of atoms as
\begin{equation}
    n{l_j}\left( F =f\right),
\end{equation}
where $j=l+s$ is the total electron angular momentum defined by electronic orbit-spin interactions, which determines the fine structure. $F=I\pm j$ is the total atomic angular momentum that governs the hyperfine structure, arising from the electron-nuclear interactions with $I$ defined as the nuclear spin\cite{haken2005physics}. Notably, in the absence of an external electromagnetic field, states that share the same $n, l, j$ are degenerate, which means that they possess identical energy. TABLE \ref{tab:quantum_numbers} summarizes the aforementioned quantum numbers, and Fig. \ref{fig:atom} illustrates the governing rules of these quantum numbers to atomic properties based on the visualized electron cloud model.

Rydberg atoms are defined as a class of atoms with high principal quantum number $n$, typically with $n>20$ \cite{cui2024rydberg,gallagher1988rydberg}. Alkali metal atoms, e.g., rubidium (Rb) and cesium (Cs), are predominantly employed in Rydberg atom research, as their hydrogenic structure facilitates both theoretical analysis and experimental validation \cite{schlossberger2024rydberg,holloway2021multiple,gallagher1988rydberg}.
The large $n$ endows Rydberg atoms with some key properties: 
\begin{itemize}
    \item \textit{Large Atomic Radius}: The orbital radius of electron follows $r\propto n^2$ when excited to a high principal quantum number state \cite{cui2024rydberg,gong2025rydberg,chen2025new}, as shown in Fig.~\ref{fig:atom}.
    
    \item \textit{High Sensitivity to Electromagnetic Fields}: Large atomic radius leads to weakened Coulomb force, rendering Rydberg atoms highly susceptible to external electromagnetic fields. This is characterized by the dipole moment $\mu$, which approximately follows $\mu\propto n^2$ \cite{zhang2024rydberg,cui2024rydberg,gong2025rydberg}.
    %, thereby enabling an exaggerated response to external fields.
    
    \item \textit{Dense Energy Levels}: The energy spacing between adjacent levels can be described by $\Delta E\propto 1/n^2-1/(n+1)^2\approx2/n^3$ \cite{liu2023electric,cui2024rydberg,zhang2024rydberg}, which decreases rapidly as $n$ increases. This results in a dense spectrum of transition lines spanning from the microwave frequency bands to terahertz frequency bands.
    
    \item \textit{Long Radiative Lifetime}: Based on the Bohr correspondence principle, Rydberg atoms with high $n$ exhibit classical physical behavior. Their radiative lifetimes scales approximately as $\tau\propto n^{3}$, which is significantly longer than the typical $ns$ lifetimes of lower excited states \cite{liu2023electric}.
    %, facilitating the maintenance of quantum coherence.
\end{itemize}

These distinctive properties constitute the physical basis for utilizing Rydberg atoms to detect external electromagnetic fields, as will be detailed in the subsequent subsections.

\begin{table}[!t]
	\centering
	\caption{Summary of Quantum Numbers and Angular Momenta.}
	\label{tab:quantum_numbers}
	\renewcommand{\arraystretch}{1.2}
	\setlength{\tabcolsep}{3pt} % 微调列间距节省空间
	\begin{tabularx}{\textwidth}{l|c|c|>{\raggedright\arraybackslash\hsize=0.9\hsize}X|>{\raggedright\arraybackslash\hsize=1.1\hsize}X}
		\Xhline{0.8pt}
		\textbf{Category}
		& \textbf{Symbol} 
		& \textbf{Value/Range} 
		& \textbf{Physical Description} 
		& \textbf{Main Impacts on Atom-Field Interaction} \\
		\Xhline{0.6pt}
		Principal & $n$ & $1, 2, 3, \cdots$ 
		& Defines electron shell energy and radial distance
		& \multirow{7}{=}{
			\begin{itemize}[leftmargin=1em, nosep, topsep=0pt, partopsep=0pt, labelsep=0.3em]
				\item \textbf{$n,l,j,F$}: Jointly determine atomic energy eigenvalues.
				\item \textbf{$n$}: Determines response sensitivity, resonant frequency and coherence time.
				\item \textbf{$l$}: Governs electric dipole transition allowance via the selection rule.
				\item \textbf{$m$}: Determines polarization matching with fields and induces Zeeman shift.
			\end{itemize}
		} \\
		\cline{1-4}
		Orbital & $l$ & $0, 1, \cdots, n-1$ 
		& Defines orbital shape and angular momentum
		&  \\
		\cline{1-4}
		Magnetic & $m$ & $0, \pm1, \cdots, \pm l$ 
		& Projection of orbital angular momentum
		&  \\
		\cline{1-4}
		Spin & $s$ & $\pm 1/2$ 
		& Intrinsic electron spin angular momentum
		&  \\
		\cline{1-4}
		\makecell[l]{Total electronic \\ angular momentum} & $j$ & $|l-s|, \cdots, l+s$ 
		& Spin-orbit coupling that determines fine structure
		&  \\
		\cline{1-4}
		Nuclear spin & $I$ & Isotope dependent 
		& Intrinsic angular momentum of the nucleus
		&  \\
		\cline{1-4}
		\makecell[l]{Total atomic \\ angular momentum} & $F$ & $|j-I|, \cdots, j+I$ 
		& Defines atomic hyperfine structure
		&  \\
		\Xhline{0.8pt}
	\end{tabularx}
\end{table}

\subsection{Atom-Field Interaction Mechanisms}\label{sec:atom-field interaction}
The internal structure of an atom defines its static energy levels, while the dynamic evolution of its quantum state is governed by interactions with external electromagnetic fields. 
Through the absorption or emission of the incident photons, atomic transition is induced between discrete stationary energy states. The energy difference between initial state $|i\rangle$ and target state $|j\rangle$ is determined by the relation $\Delta E = \hbar\omega_{ij}$, where $\hbar$ is the reduced Planck's constant and $\omega_{ij}$ denotes the atomic transition frequency. 
Denote the angular frequency of the external field as $\omega$. When $\omega = \omega_{ij}$, the external field efficiently couples the two atomic states, and the interaction strength between the external field and the atom is maximized. In practical physical systems, however, a deviation often exists between $\omega$ and $\omega_{ij}$, which is defined as the detuning $\Delta = \omega - \omega_{ij}$.

The interaction strength between atom and external field is quantified by the \textit{Rabi frequency}, defined as $\Omega = \frac{\langle i|\bm{\mu}\cdot\bm{E}|j\rangle}{\hbar}$. Here, $\bm{\mu}$ represents the transition dipole moment between the initial and final states, characterizing the intrinsic coupling strength of the atomic transition, while $\bm{E}$ denotes the amplitude of the external field. Physically, the product $\langle i|\bm{\mu}\cdot\bm{E}|j\rangle$ corresponds to the interaction energy scale, and dividing this product by $\hbar$ yields the Rabi frequency with dimensions of angular frequency. The Rabi frequency describes the frequency of periodic oscillations in the atomic population between two quantum states during time evolution. Such oscillations are known as \textit{Rabi oscillations}, which represent one of the most typical quantum dynamical behaviors observed during coherent atom-field interactions.

These interaction dynamics are fundamentally driven by the energy structure of the atomic system, 
whose foundation is the total energy of an atom represented by the \textit{Hamiltonian}, i.e.,
\begin{equation}
	\bm{H}=\underbrace{\sum_{i}\bm{E}_{i}|i\rangle \langle i|}_{\bm{H}_0}+\underbrace{\sum_{i}\sum_{j}\bm{V}_{ij}|i\rangle \langle j|}_{\bm{H}_I}, 
\end{equation}
where $\bm{H}_0$ represents the unperturbed atomic Hamiltonian with eigen energy $\bm{E}_i$, and $\bm{H}_I$ is the interaction Hamiltonian, with $\bm{V}_{ij}$ denoting the coupling strength between $|i\rangle$ and $|j\rangle$. In matrix form, $\bm{H}_0$ and $\bm{H}_I$ constitute the diagonal and off-diagonal terms of $\bm{H}$, respectively.

\subsection{Atomic Quantum Interference Phenomena}
\subsubsection{Electromagnetically Induced Transparency (EIT)}\label{sec:EIT}
Typically, atomic transitions rely on the absorption of incident electromagnetic waves, while an unique phenomenon known as \textit{electromagnetically induced transparency} (EIT) suppresses this absorption, rendering the atomic medium transparent to the waves. 
This represents a profound quantum interference phenomenon emerging from atom-field interactions, which underpins the operation of RAQ receivers. 

Building on the Hamiltonian for energy representation, we consider a typical three-level system, comprising a ground state $|g\rangle$, an intermediate excited state $|e\rangle$, and a Rydberg state $|r\rangle$, with corresponding eigen energies $\hbar\omega_g$, $\hbar\omega_e$, and $\hbar\omega_r$, respectively. A weak probe laser with amplitude $\bm{E}_p$ and frequency $\omega_p$ drives the $|g\rangle \leftrightarrow |e\rangle$ transition, while a strong coupling laser with amplitude $\bm{E}_c$ and frequency $\omega_c$ drives the $|e\rangle \leftrightarrow |r\rangle$ transition.
Under the rotating-wave approximation (RWA), the time-dependent oscillatory terms are eliminated. Then, in the interaction picture, the effective Hamiltonian governing the system dynamics is given by \cite{khan2016role}
\begin{equation}
	\hat{\bm{H}} = -\frac{\hbar}{2}
	\begin{pmatrix}
		0 & \Omega_p & 0 \\
		\Omega_p & -2\Delta_p & \Omega_c \\
		0 & \Omega_c & -2(\Delta_p + \Delta_c)
	\end{pmatrix},
	\label{eq:Hamiltonian}
\end{equation}
where $\Omega_p = \bm{\mu}_{ge}\cdot\bm{E}_p/\hbar$ and $\Omega_c = \bm{\mu}_{er}\cdot\bm{E}_c/\hbar$ are the Rabi frequencies of the probe and coupling lasers, respectively. The terms $\Delta_p = \omega_p - (\omega_e - \omega_g)$ and $\Delta_c = \omega_c - (\omega_r - \omega_e)$ represent their corresponding detuning.

By solving the eigenvalue equation $\hat{\bm{H}}|\psi\rangle = \lambda|\psi\rangle$, we can identify a unique eigen state with eigenvalue $\lambda=0$ under the two-photon resonance condition $\Delta_p + \Delta_c = 0$. This state is defined as the \textit{dark state}, expressed as
\begin{equation}
	|D\rangle = \frac{\Omega_c}{\sqrt{\Omega_p^2 + \Omega_c^2}}|g\rangle - \frac{\Omega_p}{\sqrt{\Omega_p^2 + \Omega_c^2}}|r\rangle,
	\label{eq:DarkState}
\end{equation}
which notably contains no component of the intermediate state $|e\rangle$. Physically, this absence results from destructive quantum interference between the two excitation pathways driven by the probe and coupling lasers. Consequently, the atomic medium becomes transparent to the probe laser, forming a narrow transmission window——a phenomenon known as EIT \cite{fleischhauer2005electromagnetically}. 

To form the EIT, two transition paths must be chosen that share a common energy level, each path driven by the probe laser and coupling laser, respectively. 
The stepwise excitation scheme employed in the preceding analysis exemplifies the ladder-type (or $\Xi$-type) configuration. Beyond this, there exist two alternative schemes, i.e., the V-type and $\Lambda$-type configurations, as illustrated in Fig.~\ref{fig: energy level}.

Among these, the $\Xi$-type configuration is by far the most prevalent scheme in RAQ receivers. Its stepwise energy level arrangement enables efficient population of highly-excited Rydberg states through the two-photon excitation process. Due to the huge dipole moment of the top Rydberg state, it exhibits exceptional sensitivity to external fields, making this configuration an ideal platform for signal reception.

In contrast, V-type and $\Lambda$-type configurations, while prominent in quantum optics, are currently less used for signal reception in RAQ-based systems. The V-type configuration, which comprises two excited states sharing a common ground state, is widely utilized in quantum beat spectroscopy \cite{khan2016role}. However, optical pumping tends to deplete the ground state population, which undermines stable quantum interference and complicates signal reception \cite{khan2016role}. The $\Lambda$-type configuration, where two lower states are coupled via a common excited state, excels in applications such as quantum memory and coherent population trapping \cite{lei2022eitmemory}. Yet, due to the absence of a direct Rydberg state, it is generally not the preferred choice for RAQ receivers. Therefore, the subsequent analysis is predominantly based on the $\Xi$-type configuration.

\subsubsection{Autler-Townes (AT) Splitting} \label{sec:ATS}
%On the basis of EIT configuration, the detection of external RF signal fields can be accomplished through the perturbation on its established quantum interference.
On the basis of an established EIT effect, we then introduce another important quantum interference phenomenon, the perturbation of which can be leveraged to detect external RF signal.

In addition to the probe and coupling lasers, we introduce an external RF signal with field strength $\bm{E}_s$ to the atoms, which couples $|r\rangle$ to an adjacent Rydberg state $|r'\rangle$. This extends the model to a four-level system, and the corresponding time-independent Hamiltonian can be similarly derived as
\begin{equation}
	\tilde{\bm{H}} \!=\! -\frac{\hbar}{2}
	\begin{pmatrix}
		0 & \Omega_p & 0 & 0\\
		\Omega_p & -2\Delta_p & \Omega_c & 0\\
		0 & \Omega_c & -2(\Delta_p \!+\! \Delta_c) & \Omega_s\\
		0 & 0 & \Omega_s & -2(\Delta_p \!+\! \Delta_c \!+\! \Delta_s)
	\end{pmatrix},
	\label{eq:Hamiltonian}
\end{equation}
where $\Omega_s = \bm{\mu}_{s}\cdot\bm{E}_s/\hbar$ and $\Delta_s$ denotes the Rabi frequency and detuning of the external RF signal, respectively. 
By isolating the dynamics between the two Rydberg states, we solve the eigenvalue problem for the coupling Hamiltonian
\begin{equation}
	\tilde{\bm{H}}_{\rm coup} = -\frac{\hbar}{2}
	\begin{pmatrix}
		-2(\Delta_p + \Delta_c) & \Omega_s\\
		\Omega_s & -2(\Delta_p + \Delta_c + \Delta_s)
	\end{pmatrix}.
	\label{eq:coupling Hamiltonian}
\end{equation}
Solving the eigenvalue equation $\tilde{\bm{H}}_{\rm coup}|\psi\rangle = \lambda|\psi\rangle$ yields two new eigenvalues, i.e., $\lambda _{+,-}=\frac{\hbar}{2}\left(-\Delta_{s}\pm\sqrt{\Delta_{s}^2+\Omega_{s}^2}\right)$. The corresponding eigen states represent coherent superpositions of the bare Rydberg states $|r\rangle$ and $|r'\rangle$, which are physically referred to as the \textit{dressed states}.

\begin{figure*}[!t]
	\centering
	\includegraphics[width=\linewidth]{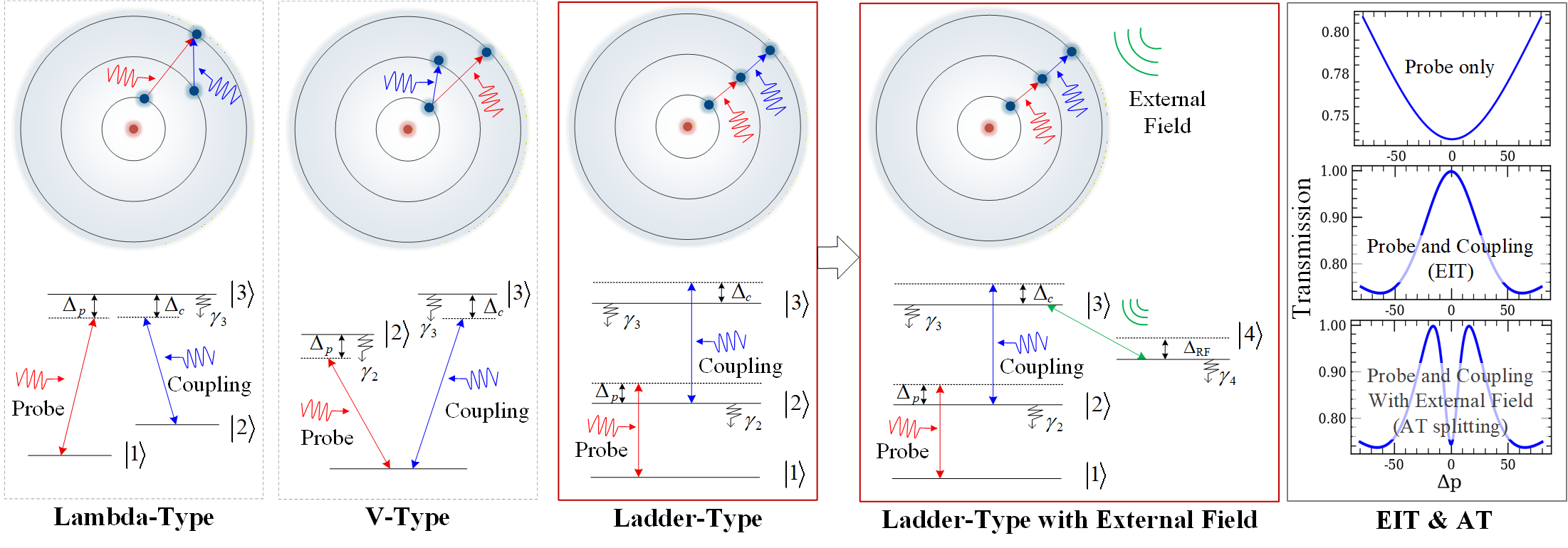}
	\caption{Left: Typical energy level diagram of EIT in three-level system, i.e., Lambda-type, V-type and Ladder-type. Right: Quantum interference phenomena under an external field, by taking the commonly used Ladder-type as example.}%. 
	\label{fig: energy level}
\end{figure*}

The presence of these dressed states perturbs the EIT dark state, disrupting the original destructive interference condition and establishing new interference pathways. This leads to the splitting of the EIT transmission peak into two distinct resonances——a phenomenon termed \textit{Autler-Townes (AT) splitting}. The splitting interval, $\Delta f_{\rm AT}$, corresponds to the energy difference between the two dressed states, given by the eigenvalues difference $\hbar\sqrt{\Delta_{s}^2+\Omega_{s}^2}$. It is proportional to the Rabi frequency of the incident RF signal \cite{liu2023electric} under resonance condition $\Delta_{s}=0$, thereby enabling the direct optical detection of RF signal through spectroscopy. Fig.~\ref{fig: energy level} (right) illustrates the EIT and AT spectra under varying conditions.

\subsection{Environmental Impacts on Atomic Systems}\label{sec:environmental-impacts}
Theoretically, atomic transitions are monochromatic, manifesting as delta-function peaks at the resonance frequency. In practice, however, absorption profiles exhibit finite widths and complex structures, which primarily stem from spontaneous emission decay and external perturbations.
The atomic spontaneous emission forms natural linewidth, which is quantified by the decoherence rate: a higher decoherence rate corresponding to a shorter lifetime and, consequently, a broader spectral line.
Note that the long lifetimes of Rydberg atoms inherently mitigate this natural broadening, which facilitates the maintenance of quantum coherence necessary for reliable reception.

Beyond intrinsic broadening, external factors including laser fields, electromagnetic environments, and thermal motion can also modulate the spectral shape. Specifically,
\begin{itemize}
    \item \textit{Power broadening}: Over-coupling between intense driving lasers and atomic transitions can lead to the broadening and potential splitting of the EIT transmission window, which is fundamentally driven by dressed-state interference dynamics \cite{khan2016role}.

    \item \textit{Doppler broadening}: Due to the thermal motion of atoms, the statistical distribution of atomic velocities results in Doppler frequency shifts, which leads to inhomogeneous broadening of the transmission spectrum  \cite{fleischhauer2005electromagnetically}.

    \item \textit{Zeeman effect}: External magnetic fields lift the inherent degeneracy of atomic  magnetic sublevels, causing energy splitting proportional to the field strength that degrade the desired quantum coherence \cite{liu2026magnetically}.

    \item \textit{Stark effect}: External electric fields perturb the energy structure, inducing Stark shifts or sub-level splitting that cause unexpected spectral displacements  \cite{fleischhauer2005electromagnetically}.
\end{itemize}
These mechanisms represent the atomic response to external physical quantities. While this responsiveness forms the theoretical basis for detecting information-carrying fields using RAQ receivers, these environmental factors simultaneously act as noise sources that constrain detection performance. Therefore, characterizing and controlling spectral broadening and shifting is pivotal for achieving efficient reception.

\subsection{Time-Evolution of Open Atomic Systems}\label{sec:time-evolution}
The aforementioned time-independent eigenvalue equation yields the stationary eigenstates and their corresponding energy eigenvalues of the atomic system. However, practical response of atomic ensemble to external electromagnetic fields is intrinsically a dynamic process, which is inevitably subjected to complicated atom-environment interactions, including spontaneous emission, transit-time broadening, and collisional dephasing. These dissipative processes induce energy decay and quantum decoherence, which drives the system from a pure quantum state into a statistical mixed state. 

To rigorously describe the dynamics of an open quantum system involving dissipation and statistical mixtures, the \textit{density matrix} that contains both the population probabilities and quantum coherences among different energy levels is introduced to characterize the atomic ensemble, and is given by
\begin{equation}
    \bm{\rho} = \sum_{n} p_n |\psi_n\rangle\langle\psi_n|,
\end{equation}
where $p_n$ is the probability of the system being in state $|\psi_n\rangle$. In the matrix representation, the elements of $\bm{\rho}$ possess distinct physical meanings:
\begin{itemize}
    \item \textit{Diagonal elements} $\rho_{nn}$: Represent the \textit{populations}, quantifying the probability of the atom being in state $|n\rangle$.
    \item \textit{Off-diagonal elements} $\rho_{nm}, n \neq m$: Represent the \textit{coherences}, characterizing the quantum interference between states $|n\rangle$ and $|m\rangle$. The magnitude of these terms reflects the degree of quantum coherence, while the decoherence caused by environmental interaction suppresses these terms over time, driving the system towards an incoherent statistical mixture.
\end{itemize}

The time evolution of $\bm{\rho}$ is governed by the Liouville equation $\dot{\bm{\rho}} = -i[\bm{H}, \bm{\rho}]/\hbar $ \cite{brasil2013simple}, which extends to the \textit{Lindblad master equation} by introducing a dissipation term $\mathcal{L}[\bm{\rho}]$, i.e.,
\begin{equation}
    \dot{\bm{\rho}} = -\frac{i}{\hbar}[\bm{H}, \bm{\rho}] + \mathcal{L}[\bm{\rho}].
    \label{eq:MasterEq}
\end{equation}
Here, the first term $-i[\bm{H}, \bm{\rho}]/\hbar$ describes the unitary evolution driven by the coherent Hamiltonian. The second term $\mathcal{L}[\bm{\rho}]$ is the Lindblad superoperator that models non-unitary dissipative processes, which is generally expressed as
\begin{equation}
    \mathcal{L}[\bm{\rho}] = \sum_{k} \gamma_k \left( L_k \bm{\rho} L_k^\dagger - \frac{1}{2} L_k^\dagger L_k \bm{\rho} - \frac{1}{2}  \bm{\rho}L_k^\dagger L_k \right),
    %\mathcal{L}[\bm{\rho}] = \sum_{k} \gamma_k \left( L_k \bm{\rho} L_k^\dagger - \frac{1}{2} \{L_k^\dagger L_k, \bm{\rho}\}\right),
\end{equation}
where $L_k$ denotes the jump operator corresponding to the $k$-th dissipation channels, e.g., spontaneous decay from $|e\rangle$ to $|g\rangle$ and dephasing of the Rydberg state, and $\gamma_k$ denotes the corresponding relaxation rates.

To solve the system dynamics numerically, the master equation in \eqref{eq:MasterEq} is projected onto the atomic energy basis states. This projection transforms the single operator equation into a set of coupled ordinary differential equations known as the \textit{Optical Bloch Equations} (OBEs), which can be generally represented as
%The OBEs describe the time-evolution of each density matrix element $\dot{\bm{\rho}}_{nm}$. Physically, these equations decompose the dynamics into two competing contributions:
\begin{equation}
    \dot{{\rho}}_{nn} = \underbrace{\sum_{E_k > E_n} \Gamma_{kn} \rho_{kk} - \Gamma_n \rho_{nn}}_{\text{Incoherent Decay}} - \underbrace{\frac{i}{2} \sum_k (\Omega_{nk} \rho_{kn} - \rho_{nk} \Omega_{kn})}_{\text{Coherent Driving}},
    \label{eq:OBE_Population}
\end{equation}
\begin{equation}
    \dot{{\rho}}_{nm} = -\left( i \Delta_{nm} + \gamma_{nm} \right) \rho_{nm} + \frac{i}{2} \sum_k \left( \Omega_{nk} \rho_{km} - \rho_{nk} \Omega_{km} \right).
    \label{eq:OBE_Coherence}
\end{equation}
Here, $\Gamma_{kn}$ denotes the spontaneous decay rate from a higher state $|k\rangle$ to $|n\rangle$, and $\Gamma_n = \sum_{E_n > E_j} \Gamma_{nj}$ is the total population decay rate out of state $|n\rangle$. The coherent atom-field interaction is characterized by the effective Rabi frequency $\Omega_{nk} = \bm{\mu}_{nk}\cdot\bm{E}/\hbar$, which is non-zero when an external field couples states $|n\rangle$ and $|k\rangle$.
$\Delta_{nm}$ represents effective detuning relative to the $|n\rangle \leftrightarrow |m\rangle$ transition. $\gamma_{nm} = (\Gamma_n + \Gamma_m)/2 + \gamma_{nm}^{\mathrm{col}}$ is the total dephasing rate, incorporating both intrinsic radiative decay and pure state interruption such as collisional broadening $\gamma_{nm}^{\mathrm{col}}$.

By solving the OBEs under the steady state condition, i.e., $\dot{\bm{\rho}} = 0$, one can derive the absorption and transmission spectra of the probe laser, thereby reproducing the EIT and AT splitting spectra.
These spectra establish the physical basis for inversely extracting the modulated information of incident electromagnetic fields, which lays the theoretical foundation of RAQ receivers discussed in the following sections.

\section{RAQ Transceiver System: Architecture and Response Model}\label{sec:RAQ Transceiver System}
This section introduces the RAQ transceiver system, including a detailed elaboration of RAQ receiver with its compositions, typical implementation architectures and signal readout methodologies, as well as the advanced atomic transmitter. On this basis, we present a general response model derived from the OBEs, which maps the atomic dynamics to optical readout and further to the mathematical input-output relationship. Finally, we review the prevalent solution frameworks for the steady-state and transient responses, and summarize the noise model by identifying the major noise sources in each component.

\begin{figure*}[!t]
	\centering
	\begin{minipage}[t]{\linewidth}
		\centering
		\includegraphics[width=\textwidth]{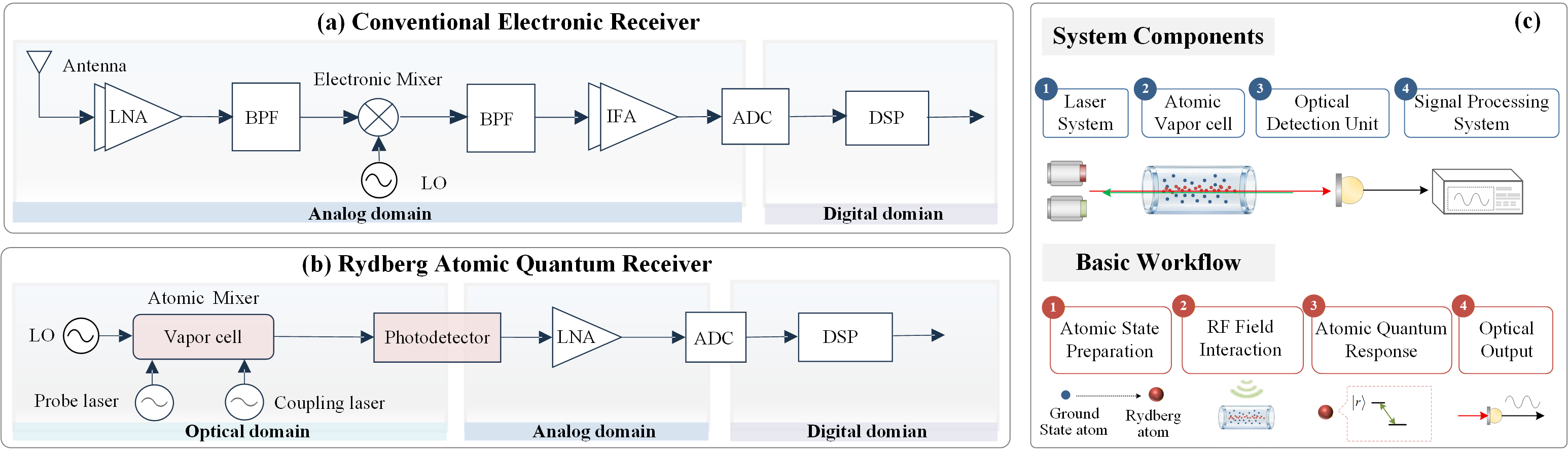}
		\caption{Comparison of (a) conventional electronic receiver and (b) RAQ receiver. (c) The system components and basic workflow of RAQ receiver. LNA: Low-noise amplifier, BPF: Band-pass filter, IFA: Intermediate frequency amplifier, ADC: Analog-to-digital converter, DSP: Digital signal processing.}
		\label{fig: comparison and workflow}
	\end{minipage}%
	\\ \vspace{5pt}
	\begin{minipage}[t]{\linewidth}
		\centering
		\includegraphics[width=\textwidth]{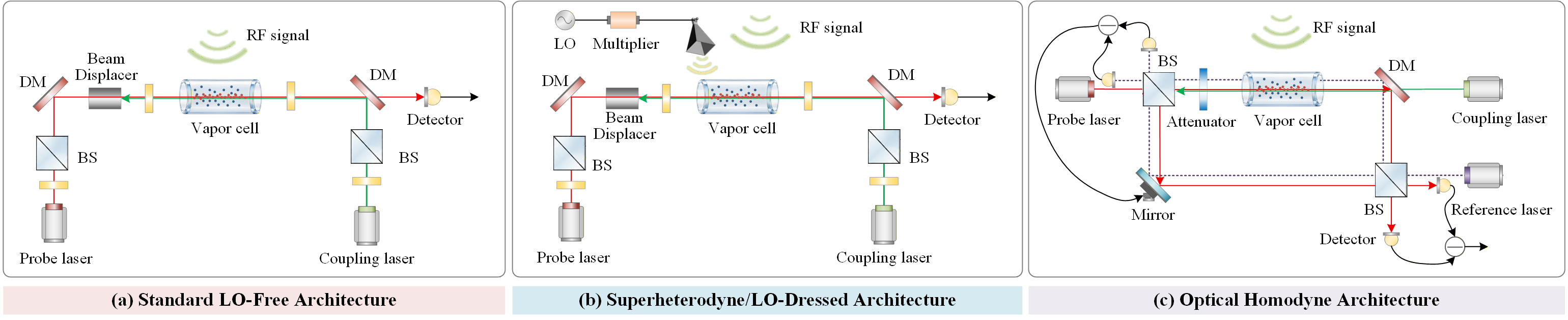}
		\caption{Typical implementation architectures of RAQ receiver. (a) Standard LO-free architecture \cite{artusio2022modern,schlossberger2024rydberg}. (b) Superheterodyne/LO-dressed architecture \cite{jing2020atomic,schlossberger2024rydberg,simons2019rydberg,artusio2022modern,wu2023linear,chen2025harnessing}. (c) Optical homodyne architecture \cite{wu2024atomic,kumar2017atom,wu2024atomic}. DM: Dichroic mirror, BS: Beam splitter.}
		\label{fig: receiver architecture}
	\end{minipage}
\end{figure*}

\subsection{System Architecture}
RAQ transceiver system primarily consists of an atomic vapor cell, laser systems, a photodetector, and signal processing devices, as depicted in Fig.~\ref{fig: comparison and workflow}. Currently, most RAQ-based systems employ a hybrid architecture, using classical electronic antennas for signal transmission and atomic systems for reception. In this configuration, the vapor cell and laser system functionally replace traditional RF front-end components (e.g., antennas, mixers, and filters), while the photodetector substitutes for the intermediate frequency (IF) amplifier and demodulator interface, bridging the optical and electrical domains  \cite{holloway2021multiple}.

\subsubsection{Receiver Structure}
RAQ receiver acts as the most mature and extensively researched component of current RAQ radio system. As illustrated in Fig. \ref{fig: comparison and workflow}c, a typical RAQ receiver primarily comprises four core components, i.e.,
\begin{itemize}
    \item \textit{Laser System}: Responsible for generating stable laser beams, which typically consists of a probe laser driving the ground state atomic transition and a coupling laser exciting the atoms to the Rydberg state.
    \item \textit{Atomic Vapor Cell}: Serving as the core sensing medium, it encapsulates an alkali atomic ensemble, e.g., Rb or Cs, that directly interacts with the incident RF field.
    \item \textit{Optical Detection Unit}: Comprises a photodetector and associated optical components, which measure the probe laser transmitted through the vapor cell and translate the optical readout into electrical signals.
    \item \textit{Signal Processing System}: Performs amplification, filtering, and demodulation on the output electrical signals to recover target parameters (e.g., amplitude, frequency and phase) and extract the encoded baseband information. 
\end{itemize}	

As illustrated in Fig. \ref{fig: comparison and workflow}c, the operational workflow of a RAQ receiver can be summarized into four sequential stages, i.e., atomic state preparation, RF field interaction, atomic quantum response, and optical output. First, the probe and coupling lasers are counter-propagated through the vapor cell, establishing an EIT quantum coherent state, which defines the operating point for signal reception. Subsequently, the incident RF electric field strongly couples with the Rydberg atoms. This interaction perturbs the atomic quantum states, disrupting or modifying the pre-existing EIT interference conditions. Specifically, the RF frequency determines which adjacent Rydberg state is resonantly coupled, while the RF field strength determines the magnitude of the induced AT splitting. Finally, variations of the microscopic quantum state map onto the macroscopic transmission spectrum of the probe laser. The photodetector captures the fluctuations in optical intensity and converts them into the electrical signal, thereby realizing the continuous reception and recovery of the incident RF signal.

Building on the basic workflow, existing RAQ receiver implementations have diversified and can be classified along two aspects, i.e., front-end conversion architectures and signal readout methodology, which are detailed below.

Regarding the front-end conversion architecture, it defines the core hardware topology of the receiver and the physical coupling mechanism between the incident RF fields and the atomic ensemble, thereby setting the fundamental performance upper bound of the system.
%Driven by the continuous pursuit of higher sensitivity and signal recovery, the implementations of Rydberg receivers have evolved from simple direct-coupling configurations to sophisticated multi-field mixing and optical interferometric systems. Currently, they can be primarily classified into three major architectures:
According to whether an additional reference field is introduced to enable  coherent reception, existing mainstream front-end architectures can be divided into incoherent direct detection (Fig.~\ref{fig: receiver architecture}a) and coherent superheterodyne or homodyne detection regimes (Fig.~\ref{fig: receiver architecture}b,c).

\begin{itemize}
	\item \textit{Standard LO-Free Architecture}: 
	As the most classical operating mode, this architecture realizes RF reception through the direct coherent coupling between the incident RF signal and the Rydberg atomic transition. Once the EIT  is established, the RF field modifies the atomic coherence and maps its information onto the probe transmission spectrum. It therefore provides a direct RF-to-optical transduction pathway, in which the atomic medium itself serves as the field-sensitive element. Due to its simple operating principle and clear spectral response, it has been widely adopted as the baseline scheme for RAQ-based signal reception, while its capabilities for coherent and ultra-weak signal reception are limited \cite{schlossberger2024rydberg,artusio2022modern}.
	
	\item \textit{Superheterodyne/LO-Dressed Architecture}:  %Designed for coherent reception and weak signal environments,  
	This architecture introduces an additional strong local oscillator (LO) field as a phase reference signal  \cite{jing2020atomic,simons2019rydberg,schlossberger2024rydberg}. Under this architecture, the atomic vapor cell functions as a natural mixer, where the coherent superposition of the LO field and the weak incident RF signal generates a beat-note component at their intermediate frequency (IF)  \cite{jing2020atomic,simons2019rydberg,artusio2022modern}. This beat-note modulates the transmission of probe laser, enabling simultaneous recovery of both amplitude and phase of the target signal. While the conversion gain provided by the strong LO field makes this architecture particularly advantageous for high-sensitivity weak signal detection, the mixing process can introduce additional nonlinear distortions when the target RF signal is strong  \cite{wu2023linear,chen2025harnessing}, which compromises the ideal linear transfer relationship.
	
	\item \textit{Optical Homodyne Architecture}: This architecture introduces additional optical components to construct a optical interferometric path. By incorporating a beam splitter and a reference optical arm, it forms an interferometer structure, e.g., a Mach-Zehnder interferometer  \cite{wu2024atomic}, to process the probe laser. This optical homodyne mechanism provides significant interferometric gain, which effectively suppresses classical technical noise limit and overcomes the photon shot-noise limit  \cite{kumar2017atom,wu2024atomic}. It further pushes the sensitivity of the RAQ receiver toward the quantum projection noise limit, albeit at the expense of increased optical system complexity and stringent alignment requirements.
\end{itemize}                                                                                    
                                              
\begin{figure*}[!t]
    \centering
    \includegraphics[width=\textwidth]{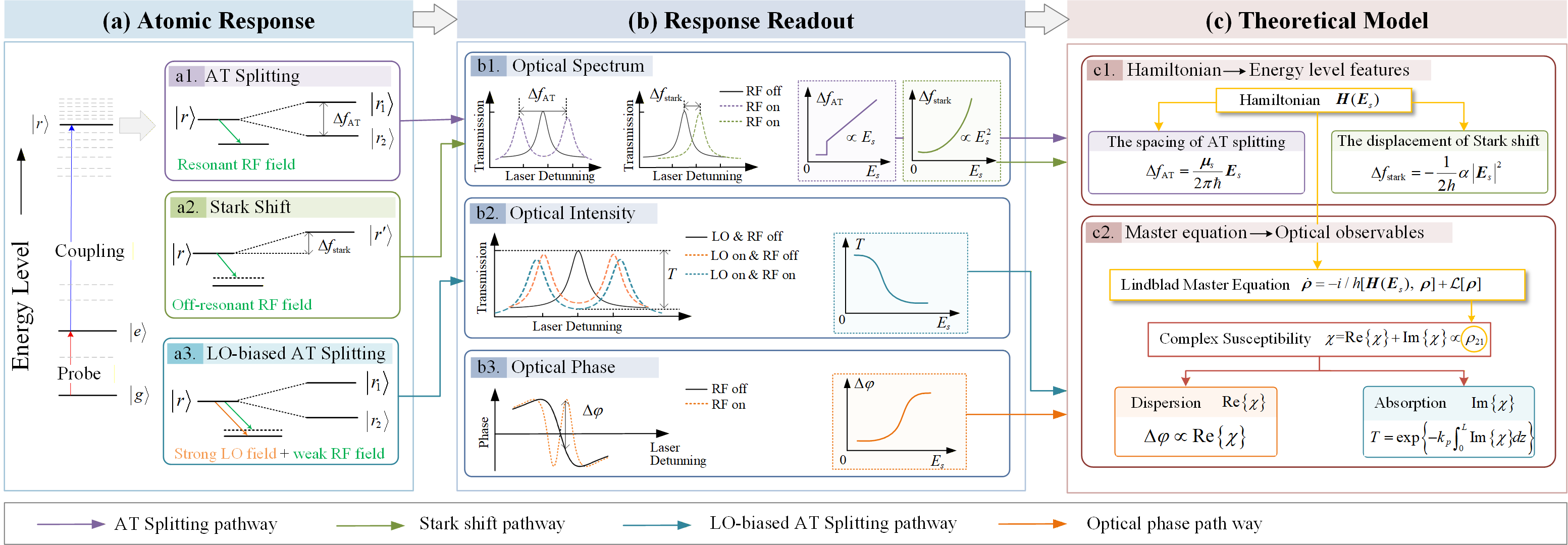}
    \caption{(a) Atomic response, (b) response readout methodologies, and (c) theoretical model of atomic response. }
    \label{Fig:Response Model}
\end{figure*}

Once the front-end architecture establishes the RF-optical mapping, the subsequent task is to convert the RF-induced atomic response into a measurable electrical signal through optical readout, where  the choice of readout method directly affects the signal extraction accuracy and acquisition speed. Based on the response feature of the EIT spectrum, the readout methodologies of RAQ receivers can be primarily categorized into the following types, as illustrated in Fig.~\ref{Fig:Response Model}b.

\begin{itemize}
	\item \textit{Optical Spectrum Measurement}: This method characterizes signal features by actively scanning the laser frequency and monitoring the RF-induced variations in the EIT transmission spectrum. The resulted spectral responses including EIT broadening, AT splitting, and Stark shifts  \cite{fancher2021rydberg}. By continuously tracking these spectral-dependent variations, information such as signal amplitude and frequency can be extracted. While this scheme provides comprehensive spectral information, the scanning process fundamentally restricts the instantaneous bandwidth and response speed of the receiver \cite{chen2025new,schlossberger2024rydberg}.
	
	\item \textit{Optical Intensity Measurement}: This method operates by locking the laser at a fixed frequency, typically at the EIT resonance center, i.e., $\Delta_p = 0$, and directly monitors the optical power of the transmitted probe laser  \cite{chen2025harnessing}. As the RF field strength increases, the resulting splitting or broadening of EIT window instantly manifests as a measurable drop in transmission  \cite{schlossberger2024rydberg}. This approach eliminates the scanning overhead, simplifies the system architecture, and enables high-speed signal acquisition, making it well-suited for detecting continuous waveforms and weak signals.
	
	\item \textit{Optical Phase Measurement}: This method detects the RF signal by monitoring the phase variations of probe laser. Governed by the Kramers-Kronig relations  \cite{yang2023machzehnder}, the RF-induced perturbation not only alters the EIT absorption or transmission spectrum, but also modifies the dispersion characteristics of the atomic vapor cell. The dispersion variation induces a proportional phase shift onto the transmitted probe laser, which can be measured to recover the information carried by the electromagnetic wave  \cite{wu2024atomic,gong2026rydberg_models}. This approach is particularly effective for coherent detection schemes where phase information is essential. 
\end{itemize}

Beneath these architectures and readout methodologies lies a unified physical mechanism, i.e., the conversion of microscopic RF-atom interactions to measurable macroscopic optical signals.
To accurately characterize this process and predict the ultimate capabilities of the receiver, it is necessary to establish a quantitative response model to bridge the external RF signal with the final optical detection. %, which will be discussed next. 
%the following section will systematically construct the quantitative response model of the receiver, mathematically bridging the external RF stimulus with the final optical detection.

\subsubsection{Transmitter Structure}
While atomic receivers dominate current research, atomic transmitters (or atomic emitters) also represent a cutting-edge frontier in quantum electromagnetic engineering. Rather than relying on classical oscillatory circuits and macroscopic metallic antennas, advanced atomic transmitters exploit controlled transitions between quantum states to generate and radiate electromagnetic signals \cite{10670667, 9905692}.  Its core components include an atomic vapor cell, a laser pumping and manipulation system, and an external field control unit.

Leveraging the physical principles of quantum superposition and entanglement, the atomic transmitter translates atomic coherence into significant signal advantages \cite{10670667}. 
\begin{itemize}
	\item \textit{Single-Photon Emission}: Enabling secure quantum communication capabilities.
	
	\item \textit{High-Order Coherence}: Emitted photons exhibit nearly identical frequency and phase, significantly enhancing frequency stability while suppressing phase noise.
	
	\item \textit{Superradiance Effect}: By utilizing the quantum correlations within atomic ensemble, the radiated power theoretically scales with the square of atom number $N^2$, instead of the linear scaling $N$ of classical systems. This enables high-power emission within a microscopic volume.
	
	\item \textit{Breaching the Chu-Harrington Limit}: It holds the potential to generate very low frequency (VLF) and extremely low frequency (ELF) radiation from electrically small sources, surpassing fundamental classical limits.
\end{itemize}

In advanced physical implementations, generating low-frequency signals requires transitions between closely spaced energy levels. This is achieved by manipulating intermediate-$n$ Rydberg atoms with DC electric fields induced Stark effect or magnetic fields induced Zeeman effect \cite{9905692,ma2017paschen,ma2022measurement}. Stimulated emission is then actively triggered and precisely controlled by external perturbations, such as ultrafast laser pulses. 

Although research on atomic transmitters remains limited, mastering this high-fidelity quantum state preparation and radiation control constitutes one of the most exciting vanguard endeavors in next-generation atomic radios.

\subsection{Atomic Response Model}
%As discussed in Section \ref{sec:RAQ Transceiver System}, the RAQ receiver extract the RF signal by monitoring probe laser passing through the vapor cell, a response model that bridge the microscopic quantum dynamics of the atomic medium with the macroscopic optical propagation is crucial. This optical response---observed experimentally as the EIT and AT splitting spectra---is fundamentally governed by the complex susceptibility, denoted as $\chi$, of the alkali atomic medium.
As discussed in Section \ref{sec:RAQ Transceiver System}, RAQ receivers extract RF information from the transmitted probe laser after its interaction with the atomic vapor cell. To describe this process, a theoretical atomic response model is required to bridge the quantum dynamics with the optical readout. 

Based on the readout methodologies summarized in Fig.~\ref{Fig:Response Model}b, the theoretical model can be established from two aspects, as shown in Fig.~\ref{Fig:Response Model}c. First, the incident RF field modifies the system Hamiltonian, thereby changing the Rydberg energy-level structure and producing observable spectral features represented by AT splitting and Stark shifts. For a resonant RF field, the spacing of AT splitting is proportional to the field strength, i.e., $\Delta f_{\rm AT}=|\bm{\mu}_{s}\cdot\bm{E}_s|/(2\pi\hbar)$, whereas for an off-resonant RF field, the Stark shift follows $\Delta f_{\rm stark}=-\alpha|\bm{E}_s|^2/(2h)$, where $\alpha$ is the dynamic polarizability at the RF frequency.

Beyond these energy-level features, the optical intensity and phase readouts are governed by the complex susceptibility $\chi$ of the atomic medium, which is determined by the density-matrix solution of the Lindblad master equation in \eqref{eq:MasterEq}. 
%in atomic coherence modifies the absorption and dispersion experienced by the probe field 
Specifically, $\chi$ provides the connection to the atomic coherence of the probe transition, i.e., $\rho_{21}$, and is defined as
\begin{equation}
	\chi = \frac{2 N_0 |\mu_{ge}|^2}{\varepsilon_0 \hbar \Omega_p} \rho_{21},
\end{equation}
where $\varepsilon_0$ denotes the vacuum permittivity, $N_0$ represents the atomic number density within the vapor cell, $\mu_{ge}$ is the transition dipole moment for the $|g\rangle \rightarrow |e\rangle$ transition, and $\Omega_p$ is the local Rabi frequency of the probe laser.

This complex coefficient encapsulates the optical properties of the atomic ensemble. The real part $\text{Re}(\chi)$ describes the optical dispersion experienced by the probe laser and provides the physical basis for optical phase readout, while the imaginary part $\text{Im}(\chi)$ governs probe absorption and therefore determines the optical intensity readout. As the probe laser propagates through the medium, its macroscopic transmission, denoted by $T$, can be modeled by integrating the position-dependent absorption over the interaction length $L$, i.e.,
\begin{equation}
	T = \frac{P_{\text{out}}}{P_{\text{in}}} = \exp\left( -k_p \int_{0}^{L} \text{Im}[\chi(z)] \, dz \right),
	\label{eq:eit_transmission}
\end{equation}
where $P_{\text{in}}$ and $P_{\text{out}}$ are the incident and output powers of probe laser, respectively, and $k_p$ is the wavenumber of the probe laser. Under uniform atomic density and driving fields, this expression reduces to the algebraic form $T \approx \exp[-k_p L \cdot \text{Im}(\chi)]$.

Consequently, establishing the quantitative mapping from the incident RF signal field to the coherence term $\rho_{21}$ is pivotal for constructing the response model of RAQ receiver. 
%Depending on whether the incident field is treated as time-invariant or dynamically modulated, this mapping can be analyzed through either the steady-state response or the transient evolution of $\rho_{21}$, both of which can be explicitly resolved by solving the OBEs introduced in Section \ref{sec:time-evolution}.
Since practical RF signals may induce either static or time-varying atomic dynamics, both the steady-state and transient responses of $\rho_{21}$ should be considered, which can be explicitly resolved by solving the OBEs introduced in Section \ref{sec:time-evolution}.

\subsubsection{Steady-State Response}
The steady-state response of RAQ receivers characterizes the equilibrium optical transmission under external fields. As the foundation of receiver modeling, it establishes the DC operating point, dynamic range, and fundamental input-output transfer function.

Mathematically, the steady-state behavior is determined by the stationary solutions of the OBEs by setting $\dot{\bm{\rho}}=0$. However, deriving an exact analytical solution for $\rho_{21}$ is exceedingly challenging in practical four-level or higher systems. The full solutions are often intractably complex, containing multiple nonlinear coupling terms between the RF and laser fields, which hinders their utility in analyzing communication system. To overcome this, researchers introduce physically motivated approximations, yielding more tractable mathematical expressions while retaining physical meaning.

The \textit{weak probe approximation} is the most widely adopted assumption in these analytical derivations \cite{chen2025harnessing, bussey2021rydberg, liao2020microwave, jia2024properties}, based on the typical EIT conditions where the probe Rabi frequency is much smaller than both the spontaneous decay rate and the intensities of other driving fields, e.g., $\Omega_p \ll \Gamma, \Omega_c$. Therefore, the atomic population is predominantly confined to the ground state, i.e., $\rho_{11} \approx 1$. This approximation effectively simplifies the complex continued fractions derived from the multi-level optical Bloch equations, eliminates high-order nonlinear absorption terms, and yields the standard EIT transmission spectrum.

For specific ladder-type configurations, researchers further exploit disparities in decay rates and coupling strengths between adjacent levels to reduce the equation order. By selectively neglecting weakly coupled density matrix elements or the virtually negligible decoherence rate of high-lying Rydberg states, e.g., assuming $\gamma_r \approx 0$, more concise analytical forms can be obtained \cite{gong2026rydberg_models}. Building upon the weak probe assumption, some studies further omit specific off-diagonal density matrix elements to derive simplified expressions that intuitively reflect the impact of the signal field on key detection metrics \cite{chen2025harnessing}. 

While these approximations successfully distill the essential physics and facilitate theoretical analysis, they inevitably cause distortion of the transmission spectrum and information loss.

\subsubsection{Transient Response}
Although the steady-state solution is commonly used for modeling,and analysis, it inherently neglects the dynamic process required to reach equilibrium. In contrast, the transient response characterizes the time evolution of the atomic system toward a new steady state following the variation of external fields. This dynamic process is governed by the time-dependent solutions of the OBEs, specifically describing the density matrix evolution where $ \dot{\bm{\rho}} \neq 0$. Compared to the steady-state solutions, analytical derivation of the transient response is far more complex. 

The temporal evolution $\rho_{21}(t)$ depends not only on intrinsic atomic parameters, but also couples strongly with the intensity and detuning of external fields. These factors jointly determine the upper limit of the system evolution speed, fundamentally constraining the response time and instantaneous bandwidth.

To elucidate these dynamics, researchers first investigated their physical origins. For instance, studies in \cite{ bohaichuk2022origins} identified that the dominant timescale of the transient response is dictated by underlying dephasing mechanisms, including transit time broadening, Rydberg-Rydberg collisions, and ionization. When the RF pulse duration approaches or falls below this settling time, traditional steady-state models become invalid, leading to measurement errors. This highlights the importance of incorporating transient effects in practical applications.

Building on this understanding, various analytical models for transient response have been developed. In \cite{ li2025analytical}, an analytical expression for probe transmission evolution is derived under the weak probe approximation and the assumption that Rydberg state decoherence is negligible compared to intermediate states. By applying Laplace transform under resonance, this solution revealed two transient mechanisms, i.e., an oscillatory decay when the coupling Rabi frequency exceeds the decoherence rate, and an overdamped monotonic settling otherwise. 
 
Furthermore, this analytical approach has been extended to superheterodyne architecture. Researchers in \cite{ren2024research} derive dynamic expressions for $\rho_{21}$ under time-varying fields to analyze amplitude attenuation and phase delay caused by transient lag at higher intermediate frequencies. From a system-level perspective, the concept of quantum transconductance is introduced in \cite{zhu2025general}. By utilizing small-signal perturbation and Laplace transforms to establish a transfer function from the input RF field to the output photocurrent, this work quantified transient relaxation as the system's impulse response and frequency bandwidth, thereby establishing a direct relationship between transient response and communication capacity.

In the frequency domain, recent investigations \cite{tang2026response} revealed the correlation between EIT linewidth and transient response speed, i.e., a narrower linewidth corresponds to an earlier roll-off of the frequency response and a slower transient response. 

%Table \ref{tab:steady state response} summarizes the approximation conditions, solution methods, and mathematical forms of the transient solutions adopted in the reviewed literature.

\subsubsection{Noise Model} \label{subsubsec:noise model}
RAQ receiver is subject to distinct noise sources from conventional receiver. The total noise from each component of RAQ receiver can be generally modeled as a signal-independent term plus a signal-dependent term, both of which follow Gaussian distribution  \cite{Wang2020,simons2018eit}, expressed as
%To facilitate modeling and analysis, most current studies simplify these complexities by modeling the noise as Gaussian-distributed \cite{Wang2020,simons2018eit}. The total noise $\bm{N}$ can be expressed as
\begin{equation}
	{N} = {Z_0}+\sqrt{{X}}{Z_1},
\end{equation} 
where \({X} \ge 0\) is the transmitted optical intensity, \({Z}_0 \sim \mathcal{N}(0,\sigma^2)\) represents signal-independent additive white Gaussian noise (AWGN) with constant variance \(\sigma^2\). \(\sqrt{{X}} {Z}_1\) represents signal-dependent AWGN with \({Z}_1 \sim \mathcal{N}(0,\varsigma^2 \sigma^2)\), where \(\varsigma^2 > 0\) is a scaling factor, typically ranging from 0 to 10.

\begin{figure}[!t]
	\centering
	\includegraphics[width=0.8\textwidth]{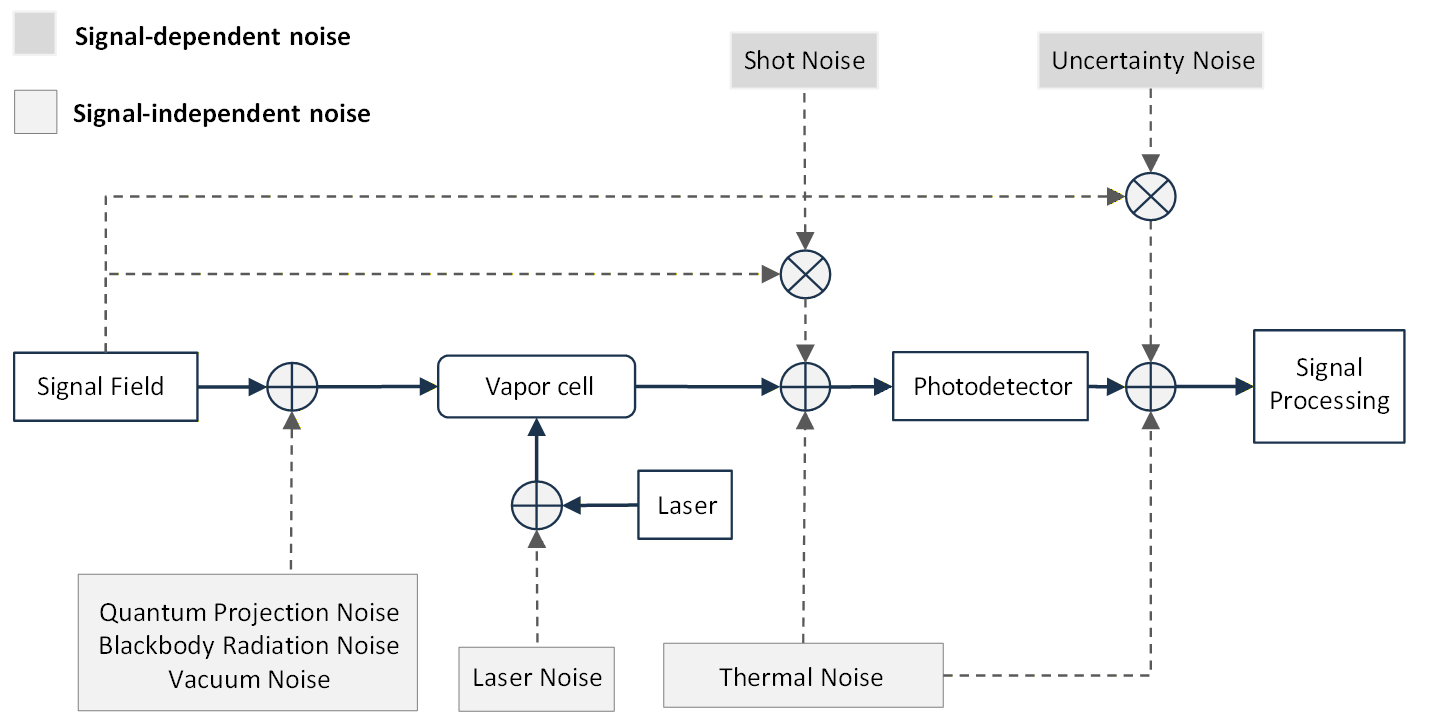}
	\caption{Noise sources and injection points in RAQ receiver chains. }
	\label{Fig:Noise Model}
\end{figure}

Signal-independent noise sources include quantum projection noise (QPN), thermal noise, blackbody radiation noise, dark current noise and laser intensity noise.
\begin{itemize}
	\item \textit{Quantum projection noise}: QPN arises from the probabilistic collapse of the wavefunction during quantum state measurement. The noise power is given by \cite{gong2026rydberg_models}
		\begin{equation}
			\sigma_{\text{QPN}}^2 = \frac{\hbar \Gamma_2 B G}{2 N t_c},
		\end{equation}
		where \(\Gamma_2\) is the decoherence rate, \(B\) is the bandwidth, \(G\) is the gain, \(N\) is the number of atoms, and \(t_c\) is the coherence time.

	\item \textit{Thermal noise}: Generated by random electron motion inside an electrical conductor, which can be expressed as \cite{chen2025harnessing}
	\begin{equation}
		\sigma_{\text{TN}}^2 = 4 k_B T B,
	\end{equation}
	with \(k_B\) Boltzmann's constant, \(T\) temperature.

	\item \textit{Blackbody radiation noise}: Arising from thermal electromagnetic radiation emitted by the environment \cite{chen2025harnessing,Botello2022}, the noise power is given by
		\begin{equation}
			\sigma_{\text{BB}}^2 = \frac{2}{\epsilon_0 c B} \cdot \frac{hf}{e^{hf/(k_B T)}-1},
		\end{equation}
		where \(h\) is Planck's constant, \(f\) is the frequency, \(\epsilon_0\) is the vacuum permittivity, and \(c\) is the speed of light.

	\item \textit{Laser noise}: Quantum fluctuations of laser field lead to random amplitude variations \cite{Tang2025}. Under the small-fluctuation approximation, its noise power is given by
	\begin{equation}
			\sigma_{{\text{LN}}}^2 \approx c^2\epsilon_0^2 S^2 E_{p}^2 \sigma_p^2,
	\end{equation}
	
	where \(E_p\) is the amplitude of the probe laser, \(\sigma_p^2\) is the variance of the random amplitude fluctuation of the probe laser, and \(S\) is the beam cross-sectional area. Although the coupling laser can also introduce noise, its contribution is sufficiently weak to be neglected because it does not directly affect the transmitted laser power \cite{Tang2025}.

\end{itemize}

Signal-dependent noise sources are dominated by various forms of shot noise and measurement uncertainty.

\begin{itemize}
	
	\item \textit{Shot noise:} Shot noise originates from three independent sources in photodetection, all arising from the random nature of carrier generation \cite{Botello2022,chen2025harnessing,gong2026rydberg_models,Wang2020}: (i) signal photon noise due to the random arrival of signal photons, (ii) background photon noise from ambient light, and (iii) dark current noise from thermal generation even in the absence of light. The total variance can be expressed as
	\begin{equation}  \sigma_{\text{shot}}^2 = i_s^2 + i_b^2 + i_d^2 = 2qB(\eta X + X_b + I_d),  \end{equation} 
	where \(i_s^2=2qB\eta X\), \(i_b^2=2qB X_b\), and \(i_d^2=2qB I_d\) denote the signal photon, the background photon, and the dark current noise components, respectively. Here, \(q\) is the electron charge, \(\eta\) is the quantum efficiency, \(X_b\) is the background light intensity, \(I_d\) is the dark current intensity.

	\item \textit{Uncertainty noise}: Observation uncertainty noise arises when weak RF signals blur the AT doublet, which can be modeled as \cite{artusio2022modern,chen2025harnessing}
	\begin{equation}
		\sigma_{\text{UN}}^2 = (\tilde{\epsilon} E_{s})^2,
	\end{equation}
	where \(\tilde{\epsilon} \approx 0.5\%\) is the fractional uncertainty and \(E_{s}\) the RF field strength.
\end{itemize}

Overall, signal-independent noise mainly determines the noise floor of the receiver, while signal-dependent noise reflects the influence of the RF field and the measurement process on system performance. %The modeling of noise connects noise sources arising from different physical mechanisms to the receiver response.
 
\begin{figure*}[!t]
	\centering
	\includegraphics[width=0.7\textwidth]{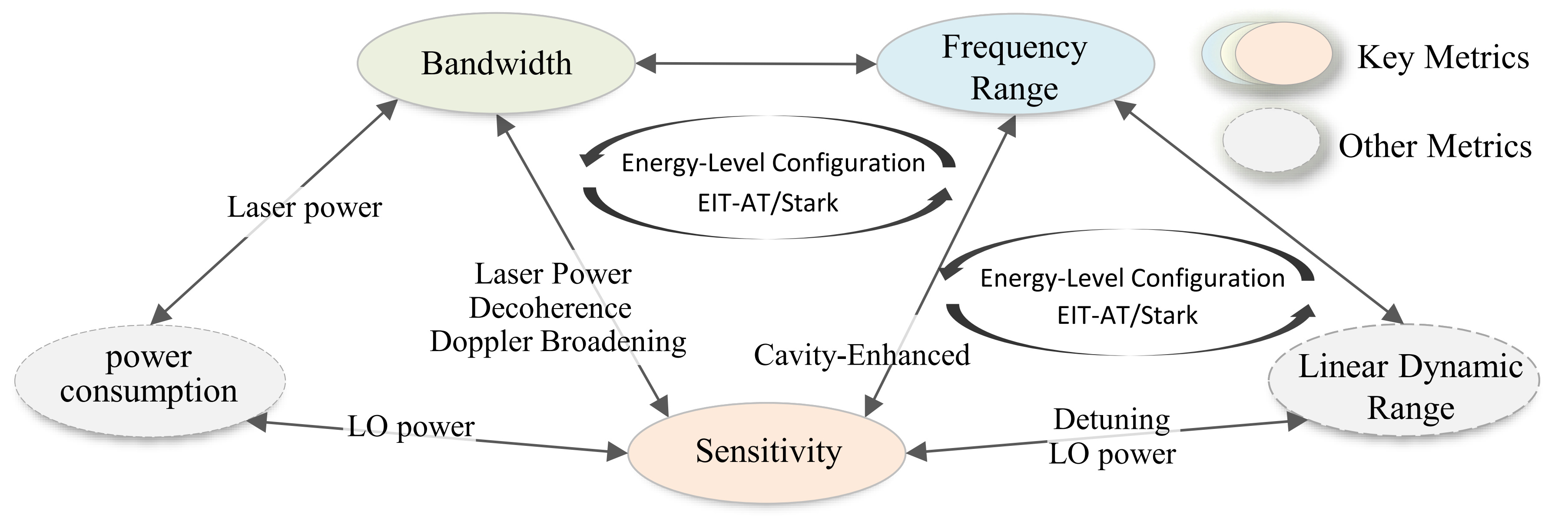}  % 设置为页面宽度的80%
	\caption{Trade-offs among key performance metrics of RAQ receivers. The black arrows denote interdependencies between different metrics. }
	\label{fig_tradeoff}
\end{figure*}

\section{Performance Trade-Off and Enhancement Techniques for RAQ Radio}
%\section{Implementation Enhancement on Rydberg Atomic Receivers}
To improve the performances of RAQ receivers, researchers have conducted extensive work. Current efforts primarily focus on three core metrics, i.e., sensitivity, instantaneous bandwidth, and operating frequency. This section analyzes the trade-offs and enhancement techniques related to these metrics, introduces their core technical approaches and innovative strategies, which provides a reference for future research in this field.

\subsection{Trade-offs Among Sensitivity, Instantaneous Bandwidth and Operating Frequency}

As key performance metrics of RAQ receivers, sensitivity, instantaneous bandwidth, and frequency range are not independently scalable, but are jointly constrained by atomic coherence and optical readout mechanisms, as shown in Fig.~\ref{fig_tradeoff}. 

The most fundamental trade-off lies between sensitivity and instantaneous bandwidth. 
%Sensitivity and instantaneous bandwidth usually constrain each other. 
A broader EIT transparency window enables larger instantaneous bandwidth, while results in reduced sensitivity. The width of the transparency window is influenced by decoherence, laser power, and Doppler effects. Therefore, improving sensitivity generally requires a narrower response window, whereas bandwidth enhancement often comes at the cost of reduced field resolution. A similar trade-off exists between sensitivity and available frequency range, where cavity-enhanced techniques can improve sensitivity but limit the accessible frequency band due to their resonant nature.

The choice between the EIT-AT effect and the Stark effect, together with the energy level configurations, further reshapes these trade-offs. The Stark effect enables off-resonant and low-frequency field detection, which expands the available frequency range. However, compared with the EIT-AT effect, it may face limitations in instantaneous bandwidth and sensitivity. Different energy level configurations offer additional degrees of freedom to relax these trade-offs, providing a promising route for simultaneous improvement of multiple metrics \cite{yang2024highly}.

Beyond three key metrics, linear dynamic range and power consumption are also important for system design. Linear dynamic range is jointly influenced by the LO operating point and detuning. Adjustment of the LO operating point enables a trade-off between sensitivity and linear dynamic range, while detuning affects the metrics by reshaping the EIT spectrum and modifying the atomic response \cite{simons2016using}. Power consumption is another key indicator, which is affected by the choices of laser power and LO power, whose choices also influence instantaneous bandwidth and sensitivity. Therefore, appropriate optical and RF power allocation is required to balance receiver performance and energy consumption.

\subsection{Sensitivity Enhancement Techniques}

Sensitivity of RAQ receivers is jointly determined by atom-field coupling strength, atomic coherence properties, optical readout efficiency, and technical noise. Studies have shown that key system parameters, e.g., frequency detuning, transit relaxation, laser driving conditions, and multi-level atomic dynamics, can affect the achievable sensitivity and should be carefully optimized \cite{wu2023theoretical,bussey2024numerical,anderson2018vapor}. Based on these considerations, sensitivity enhancement techniques can be broadly classified into passive field enhancement through resonant structures, atomic response engineering, and readout-architecture optimization. This section surveys the implementation methods and achievable performances of these approaches.

\begin{table*}[!t]
	\caption{Summary of Sensitivity Enhancement Techniques}
	\label{tab:sensitivity}
	\centering
	\renewcommand{\arraystretch}{1.3}
	\setlength{\tabcolsep}{3pt}
	\resizebox{\textwidth}{!}{
	\begin{tabular}{c|c!{\vrule width 0.6pt}c|c|c|c}
		\Xhline{0.65pt}
		\multicolumn{2}{c!{\vrule width 0.6pt}}{\textbf{Category}} & \textbf{Design Method} & \textbf{Available Frequency Range} & \textbf{Sensitivity} & \textbf{References} \\
		\Xhline{0.65pt}
		\multirow{14}{*}{\makecell[c]{Resonant\\ Cavity\\ Design}} 
		& \multirow{4}{*}{\makecell[c]{Localized\\ field\\ enhancement}} 
		& \makecell[c]{Multilayered split-ring resonator} 
		& 20.51 MHz 
		& 29.09 nV/cm/$\sqrt{\mathrm{Hz}}$ @ 20.51 MHz 
		& \cite{wan2026deepsubwavelength} \\ \cline{3-6}
		& & \makecell[c]{Photonic-crystal slot-waveguide vapor cell} 
		& 37.36 GHz 
		& 600 nV/cm/$\sqrt{\mathrm{Hz}}$ @ 37.36 GHz 
		& \cite{amarloo2025photonic} \\ \cline{3-6}
		& & \makecell[c]{Metallic waveguide resonant cavity} 
		& 14.98-15.2 GHz 
		& Field enhancement factor of 14 
		& \cite{sandidge2023field} \\ \cline{3-6}
		& & \makecell[c]{Ridge-compensated field uniformity} 
		& 9.91-10.56 GHz 
		& 1.58 $\mu$V/m/$\sqrt{\mathrm{Hz}}$ @ 10.22 GHz 
		& \cite{sandidge2024resonant} \\ \cline{2-6}
		
		& \multirow{6}{*}{\makecell[c]{Frequency-agile\\ enhancement}} 
		& \makecell[c]{Varactor-tuned PCB-enhanced resonator} 
		& 187-293.6 MHz 
		& 1.73 $\mu$V/cm/$\sqrt{\mathrm{Hz}}$ @ 250.5 MHz 
		& \cite{yang2024pcb} \\ \cline{3-6}
		& & \makecell[c]{Varactor-tuned PIN-switched resonator} 
		& 231-586 MHz 
		& 30.4 nV/cm/$\sqrt{\mathrm{Hz}}$ @ 564.2 MHz 
		& \cite{yang2026broadband} \\ \cline{3-6}
		& & \makecell[c]{Copper-column mechanical tuning} 
		& 210-308.6 MHz 
		& 179.37 nV/cm/$\sqrt{\mathrm{Hz}}$ @ 223.7 MHz 
		& \cite{li2024compact} \\ \cline{3-6}
		& & \makecell[c]{Movable-patch cage-shaped resonator} 
		& 1-2 GHz 
		& Electric field gain $>$ 40 dB 
		& \cite{tang2025frequency} \\ \cline{3-6}
		& & \makecell[c]{Tunable Fabry--Perot cavity} 
		& 4.8-12.2 GHz 
		& 5 nV/cm/$\sqrt{\mathrm{Hz}}$ @ 8.57 GHz 
		& \cite{wu2024enhancing} \\ \cline{3-6}
		& & \makecell[c]{Continuous-passband cascaded cavity} 
		& 2.781-3.008 GHz 
		& 1.47 $\mu$V/cm/$\sqrt{\mathrm{Hz}}$ 
		& \cite{gao2025cascaded} \\ \cline{2-6}
		& \multirow{4}{*}{\makecell[c]{LO-integrated\\ heterodyne\\ enhancement}} 
		& \makecell[c]{Square-spiral LOIR} 
		& 14-16 MHz 
		& / 
		& \cite{zhou2024miniaturized} \\ \cline{3-6}
		& & \makecell[c]{Square-spiral resonator with sapphire cell} 
		& 15.54 MHz 
		& 2.60 nV/cm/$\sqrt{\mathrm{Hz}}$ @ 15.54 MHz 
		& \cite{zhou2025high} \\ \cline{3-6}
		& & \makecell[c]{Standing-wave type LOIR} 
		& 264.8 MHz 
		& 104.34 nV/cm/$\sqrt{\mathrm{Hz}}$ @ 264.8 MHz 
		& \cite{yang2025local} \\ \cline{3-6}
		& & \makecell[c]{Traveling-wave type LOIR} 
		& 264.2 MHz 
		& 105.54 nV/cm/$\sqrt{\mathrm{Hz}}$ @ 264.2 MHz 
		& \cite{yang2025local} \\ \hline
		
		\Xhline{0.5pt}
		\multicolumn{2}{c!{\vrule width 0.6pt}}{\multirow{5}{*}{\makecell[c]{Atomic Response \\ Engineering}}} & \makecell[c]{Frequency comb based on pulsed lasers} & 20 kHz - 96 MHz & 2.9 $\mu$V/cm/$\sqrt{\mathrm{Hz}}$ @ 66 MHz, 88 MHz & \cite{di2025electric} \\ \cline{3-6}
		\multicolumn{2}{c!{\vrule width 0.6pt}}{} & \makecell[c]{Sideband amplification in six-wave mixing} & 16.0304 GHz $\pm$ 2 MHz & 62 nV/cm/$\sqrt{\mathrm{Hz}}$ & \cite{yang2024highly} \\ \cline{3-6}
		\multicolumn{2}{c!{\vrule width 0.6pt}}{} & \makecell[c]{Many-body enhancement } & 16.68 GHz & 49 nV/cm/$\sqrt{\mathrm{Hz}}$ @ 16.68 GHz & \cite{ding2022enhanced} \\ \cline{3-6}
		\multicolumn{2}{c!{\vrule width 0.6pt}}{} & \makecell[c]{Ground state repumping laser} & 17.0434 GHz & 0.03 $\mu$V/cm/$\sqrt{\mathrm{Hz}}$ @ 17.0434 GHz & \cite{prajapati2021enhancement} \\ \cline{3-6}
		\multicolumn{2}{c!{\vrule width 0.6pt}}{} & \makecell[c]{Collinear three-photon excitation} & 108.9 GHz & 1.25 $\mu$V/cm/$\sqrt{\mathrm{Hz}}$ @ 108.9 GHz & \cite{bohaichuk2023three} \\ \hline
		\Xhline{0.5pt}
		\multicolumn{2}{c!{\vrule width 0.6pt}}{\multirow{6}{*}{\makecell[c]{Readout and architecture \\ optimization}}} & \makecell[c]{Optical homodyne detection} & 5.047 GHz & 5 $\mu$V/cm/$\sqrt{\mathrm{Hz}}$ @ 5.047 GHz & \cite{kumar2017atom} \\ \cline{3-6}
		\multicolumn{2}{c!{\vrule width 0.6pt}}{} & \makecell[c]{Dispersion signal amplification} & 8.556-8.566015 GHz & 111 nV/cm/$\sqrt{\mathrm{Hz}}$ @ 8.566 GHz & \cite{jiang2025quantum} \\ \cline{3-6}
		\multicolumn{2}{c!{\vrule width 0.6pt}}{} & \makecell[c]{Probe laser array optimization} & 8.57 GHz & 19 nV/cm/$\sqrt{\mathrm{Hz}}$ @ 8.57 GHz & \cite{wu2024enhancingsgf} \\ \cline{3-6}
		\multicolumn{2}{c!{\vrule width 0.6pt}}{} & \makecell[c]{Heterodyne + homodyne detection + \\ detunings of laser and microwave field} & 20.64GHz & 0.185 nV/cm/$\sqrt{\mathrm{Hz}}$ @ 20.64 GHz & \cite{wu2024atomic} \\ \cline{3-6}
%		\multicolumn{2}{c!{\vrule width 0.6pt}}{} & \makecell[c]{Heterodyne reception architecture \\ frequency detuning optimization \\ transit relaxation effect correction} & / & / & \cite{wu2023theoretical} \\ \cline{3-6}
%		\multicolumn{2}{c!{\vrule width 0.6pt}}{} & \makecell[c]{Numerical model of N-level cascade system} & / & / & \cite{bussey2024numerical} \\\hline
		\Xhline{0.65pt}
	\end{tabular}
}
\end{table*}

\subsubsection{Resonant Cavity Design}
%Research on cavity design can be categorized into two types. The first type focuses on the improvement and integration of mature resonant structures, pursuing ultimate performance and miniaturization. The second type is committed to developing new resonant cavity structures to fundamentally expand the device boundaries in terms of bandwidth, tunability, and sensitivity \cite{anderson2018vapor}.
Existing studies can be categorized into three directions. The first focuses on localized field enhancement, where resonant or waveguiding concentrate the incident RF field within the atom-laser interaction region. The second pursues frequency-agile enhancement, aiming to overcome the narrowband nature of conventional resonators. The third emphasizes LO-integrated heterodyne enhancement, where the resonator is co-designed with the LO feeding path.
% \cite{anderson2018vapor}

For localized field enhancement, W. Wan \textit{et al.} \cite{wan2026deepsubwavelength} introduced a deep-subwavelength multilayered split-ring resonator (MSRR), where the staggered multilayer structure lowers the resonant frequency while maintaining strong field. H. Amarloo \textit{et al.} \cite{amarloo2025photonic} introduced a photonic crystal structure to the vapor cell, where slot-waveguide confinement and the slow-light effect jointly enhance RF-atom interaction inside the vapor cell. Furthermore, G. Sandidge \textit{et al.} \cite{sandidge2023field} emphasized that the key design objective is not only RF field amplification but also electric-field uniformity across the  interaction region. Their later work introduced narrow ridges to compensate for field decay near the vapor-cell edges, yielding a more uniform co-polarized RF field and improved sensitivity \cite{sandidge2024resonant}. 

To further enhance frequency agility, K. Yang \textit{et al.} \cite{yang2024pcb} adopted a PCB-based electrically tunable resonator (PETR), where a tuning varactor dynamically adjusts the equivalent capacitance and enables continuous VHF-band frequency tuning. They further proposed a broadband frequency-reconfigurable resonator (BFRR), extending resonator-assisted enhancement toward wider tunable frequency coverage \cite{yang2026broadband}. Similarly, Y. Li \textit{et al.} \cite{li2024compact} developed a compact tunable enhancement resonator (CTER), in which a sliding copper tuning column changes the effective current path to achieve broadband mechanical tunability. X. Tang \textit{et al.} \cite{tang2025frequency} designed a cage-shaped resonator (CSR), enabling L-band continuous tuning by moving a copper tuning patch along the resonant current loop. Based on the Fabry-Perot interference principle, B. Wu \textit{et al.} \cite{wu2024enhancing} designed a tunable open Fabry-Perot (FP) cavity that enhances the local microwave field through multiple-reflection superposition, providing high sensitivity in the C/X-band range.  Beyond point-wise tuning, G. Gao \textit{et al.} \cite{gao2025cascaded} proposed a cascaded cavity resonator (CCR) with three-stage rectangular cavity cascading and multi-cavity coupling, achieving continuous S-band field enhancement without stepwise tuning.

For LO-integrated heterodyne enhancement, A. Zhou \textit{et al.} \cite{zhou2024miniaturized} integrated an LO port with a square-spiral resonator to realize compact near-field LO coupling. In their subsequent HF-band work \cite{zhou2025high}, the square-spiral resonator was combined with a sapphire Cs vapor cell to mitigate electric-field screening. K. Yang \textit{et al.} \cite{yang2025local} developed standing- and traveling-wave LO port-integrated resonators (LOIRs) based on split-ring structures, reducing LO field attenuation and multipath interference.

\subsubsection{Atomic Response Engineering}
Inspired by the inherent physical properties of atoms, some studies focus on the design of transition paths for atom-field interactions. Sensitivity is effectively improved by optimizing the matching between transition channels and enhancing coupling efficiency.

N. Prajapati \textit{et al.} \cite{prajapati2021enhancement} introduced a ground-state repumping laser to recycle dark-state atoms back into the EIT cycle, thereby increasing the number of interacting atoms without noticeable spectral broadening and improving the EIT signal amplitude. To suppress linewidth-induced sensitivity degradation, S. M. Bohaichuk \textit{et al.} \cite{bohaichuk2023three} designed a collinear three-photon excitation scheme that matches the laser wavevectors, significantly narrowing the spectral linewidth for weak RF-field measurements. Beyond linear atomic excitation, B. Yang \textit{et al.} \cite{yang2024highly} explored six-wave mixing processes in a superheterodyne architecture, enhancing sideband generation efficiency and expanding instantaneous bandwidth while maintaining high sensitivity. D. S. Ding \textit{et al.} \cite{ding2022enhanced} further exploited nonlinear interactions and collective phase transitions in many-body systems near the critical point, where the transmission spectrum becomes highly sensitive to tiny frequency shifts, thus improving electric-field measurement accuracy. K. Di \textit{et al.} \cite{di2025electric} constructed a Rydberg atomic frequency comb by matching the pulsed-laser repetition frequency with the atomic velocity distribution, enabling broader frequency coverage and improved sensitivity without external microwave modulation.
﻿

\subsubsection{Readout and Architecture Optimization}
%\textcolor{red}{Architecture-oriented optimization focuses on the innovation in signal detection structure. It enhances sensitivity by changing the types, numbers, or combinations of receiver components.}
Readout and architecture optimization aims to improve the front-end detection topology and optical readout chain of RAQ receivers. By introducing interferometric readout, optical homodyne/heterodyne detection, probe laser arrays, weak-measurement-assisted amplification, and system-parameter optimization, these methods enhance RF-to-optical signal extraction, suppress technical noise, and improve receiver sensitivity.

S. Kumar \textit{et al.} \cite{kumar2017atom} adopted optical homodyne detection combined with a Mach--Zehnder interferometer (MZI) to suppress probe laser noise and enhance sensitivity. Following this interferometric readout route, S. Wu \textit{et al.} \cite{wu2024atomic} combined an atomic heterodyne receiver with an MZI and evaluated the received signal through laser phase variations, achieving improved sensitivity by optimizing the frequency detunings of laser and microwave fields. To further mitigate technical noise, Y. Jiang \textit{et al.} \cite{jiang2025quantum} introduced quantum weak measurement with a Sagnac loop interferometer (SLI), amplifying dispersion-induced phase differences while suppressing absorption-related disturbances. Beyond single-beam readout, B. Wu \textit{et al.} \cite{wu2024enhancingsgf} constructed optical probe laser arrays using fiber-coupled arrays and cascaded diffraction gratings, increasing the number of interacting  atoms for coherent signal accumulation and noise suppression.

A performance comparison of typical sensitivity enhancement schemes is summarized in Table \ref{tab:sensitivity}.

\subsection{Instantaneous Bandwidth Enhancement Techniques}
Instantaneous bandwidth is primarily limited by the response time of RAQ system. 
This response is governed by multiple factors, including steady-state EIT formation, transient atom-RF interaction, optical Rabi frequencies, vapor density, and technical readout bandwidth \cite{sapiro2020time}. 
For most EIT-based receivers, the dominant limitation is the formation and relaxation time of the EIT response, which is approximately on the order of 100 ns, corresponding to a very limited bandwidth of 10 MHz \cite{anderson2021atomic,sapiro2020time}.
In the following, we review advances in broadening the instantaneous bandwidth through atomic state manipulation and  multiplexed detection architectures design.

\begin{table*}[!t]
	\caption{Summary of Instantaneous Bandwidth Improvement Techniques}
	\label{tab:comparison_multirow}
	\centering
	\renewcommand{\arraystretch}{1.3}
	\setlength{\tabcolsep}{3pt}
	\resizebox{\textwidth}{!}{
	\begin{tabular}{c|c|c|c|c}
		\Xhline{0.6pt}
		\textbf{Category} & \textbf{Design Method} & \textbf{Signal Frequency} & \textbf{Instantaneous Bandwidth} & \textbf{References} \\
		\Xhline{0.6pt}
		\multirow{5}{*}{\makecell[c]{\\ Atomic State \\ Manipulation}} & \makecell[c]{Multi-dress-state engineering} & 16.03 GHz & $\Delta_n/2\pi$=12 MHz: 76.8 MHz & \cite{yan2025multi} \\ \cline{2-5}
		& \makecell[c]{Sideband amplification in six-wave mixing} & 16.0304 GHz & 10.2 MHz & \cite{yang2024highly} \\ \cline{2-5}
		& \makecell[c]{Using tunable Rydberg blockade effect} & / & 5-times larger& \cite{zhang2018fast} \\ \cline{2-5}
		& \makecell[c]{Coarse laser tuning with fine Stark shift compensation} & 2.6-3.6 GHz & GHz-level & \cite{chen2025radar} \\ \cline{2-5}
		& \makecell[c]{Zeeman effect with gradient magnetic field regulation} & 2.904 GHz & 3.2 MHz & \cite{qimeng2025instantaneous1} \\
		\Xhline{0.6pt}
		\multirow{5}{*}{\makecell[c]{\\ \\ Multiplexed Detection \\ Architectures}} & \makecell[c]{Spatiotemporal multiplexing of probe laser} & 18.14 GHz & 100 MHz & \cite{knarr2023spatiotemporal} \\ \cline{2-5}
		& \makecell[c]{Optical frequency comb probe \\ with sideband measurement on each comb teeth} & 4.0 GHz & \makecell[c]{+ sideband: 6$\pm$4 MHz\\ - sideband: 12$\pm$1 MHz} & \cite{artusio2024increased} \\ \cline{2-5}
		& \makecell[c]{Microwave-frequency-comb LO dressing} & 4.485 GHz & \makecell[c]{Per-channel 3-dB bandwidth 300 kHz; \\ instantaneous comb coverage 125 MHz} & \cite{zhang2022rydberg1} \\ \cline{2-5}
		& \makecell[c]{Stark comb with scalable vapor cell arrays} & 8.025-8.235 GHz & 210 MHz & \cite{jiao2025arbitrary} \\ \cline{2-5}
		%& \makecell[c]{Optical frequency comb-stabilized\\ millimeter-wave Rydberg reception} & 95.992512 GHz & 123.3-129.8 kHz & \cite{legaie2024millimeter} \\
		\Xhline{0.6pt}
	\end{tabular}
}
\end{table*}

\subsubsection{Atomic State Manipulation}
Manipulating atomic energy levels, quantum states, and nonlinear interaction pathways are the effective approaches to broaden the RF-to-optical response. 
Q. Zhang \textit{et al.} \cite{zhang2018fast} employed the Rydberg blockade effect to shorten the atomic response time and improve bandwidth-related characteristics at the quantum-state level. Y. Yan \textit{et al.} \cite{yan2025multi} reshaped the atomic response through energy-level dressing and the AC Stark effect, where a detuning-dependent dual-peak structure and Rabi-frequency-driven dip-lifting effect were used to broaden the detectable response range. Through gradient magnetic field manipulation, Q. Wang \textit{et al.} \cite{qimeng2025instantaneous1} realized Zeeman-sublevel-selective excitation using spatially varying magnetic fields generated by anti-Helmholtz coils, thereby expanding the instantaneous bandwidth while maintaining high sensitivity. Beyond energy-level control, B. Yang \textit{et al.} \cite{yang2024highly} optimized six-wave mixing processes in a superheterodyne architecture, enhancing negative-sideband generation efficiency and expanding the instantaneous bandwidth while preserving high sensitivity. M. Chen \textit{et al.} \cite{chen2025radar} further compensated for the discreteness of atomic energy levels through AC Stark tuning combined with laser tuning, synthesizing an equivalent broadband detection range.

\subsubsection{Multiplexed Detection Architectures}
Towards architecture design, Samuel H. Knarr \textit{et al.} \cite{knarr2023spatiotemporal} proposed a spatiotemporal multiplexing scheme that combines the pulsed probe operation and spatial division, thereby alleviating the relaxation time bottleneck of conventional single-channel architectures.To support scalable bandwidth expansion, Yuechun Jiao \textit{et al.} \cite{jiao2025arbitrary} proposed a Stark comb regulation scheme, where position-dependent Stark fields are applied to different vapor cells in an array and matched with microwave-frequency-comb spectral lines, enabling spatially parallel detection. Alexandra B. Artusio-Glimpse \textit{et al.} \cite{artusio2024increased} introduced an optical frequency comb (OFC)-based detection scheme, in which multiple probe-comb teeth perform parallel sampling and extraction of RF-induced sidebands, improving the sensitivity bandwidth through multi-channel optical detection. In addition, Li-Hua Zhang \textit{et al.} \cite{zhang2022rydberg1} developed a Rydberg microwave frequency comb spectrometer architecture. Through multi-frequency LO-field dressing and dual-microwave-frequency-comb algorithms, this architecture achieved wide-range broadband detection. %Legaie R \textit{et al.} adopted OFC-stabilized lasers and a heterodyne (HET) architecture. By optimizing 3-dB and 6-dB bandwidth selectivity metrics, this work advanced wideband detection toward millimeter-wave (MMW) applications \cite{legaie2024millimeter}.

The various enhancement techniques of instantaneous bandwidth are summarized in Table \ref{tab:comparison_multirow}.

\begin{figure*}[!t]
	\centering
	\includegraphics[width=\textwidth]{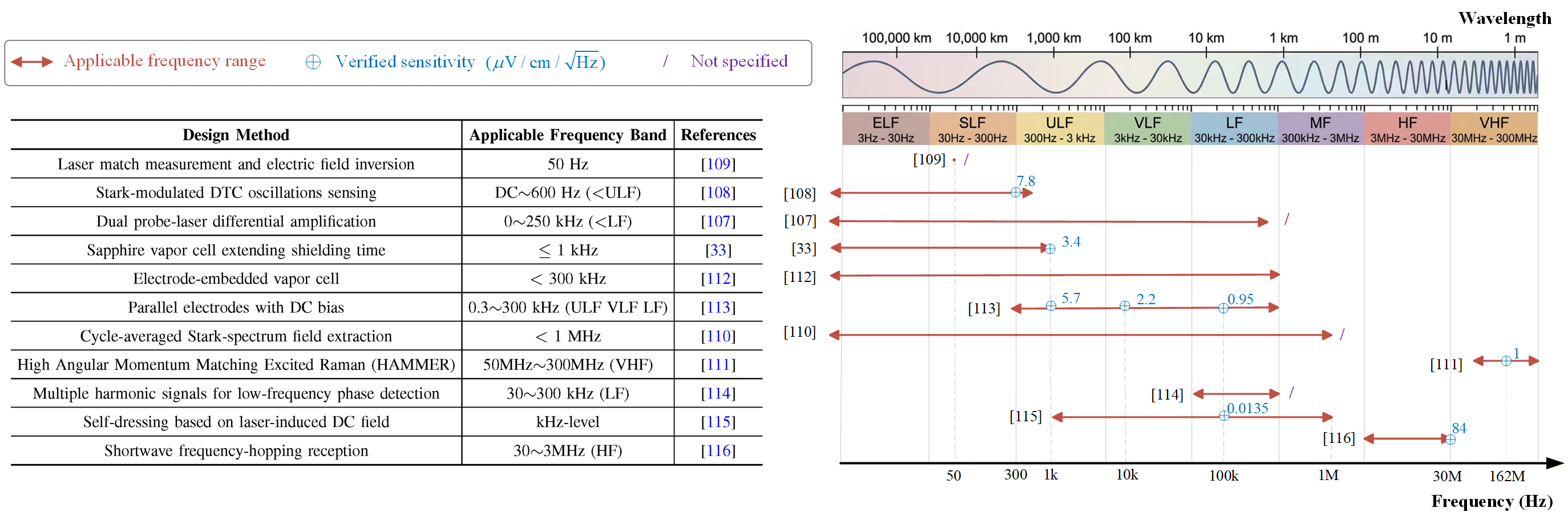} 
	\caption{Summary of operating frequency enhancement techniques, and the verified sensitivity achieved in relevant references. }
	\label{fig:Frequency range}
\end{figure*}

\subsection{Operating Frequency Enhancement Techniques}

As the target frequency decreases, resonant EIT--AT detection is constrained by the discrete spacing between adjacent Rydberg levels, since lower-frequency coupling generally requires extremely high principal quantum numbers \cite{paradis2019atomic}. Therefore, low-frequency RAQ receiver often turns to off-resonant mechanisms, especially the Stark effect, where the incident electric field is inferred from RF-induced energy-level shifts.

Specifically, Q. F. Wang \textit{et al.} \cite{wang2025measurement} combined three-photon EIT with the Stark effect and detected 250 kHz arc signals using dual probe-laser differential amplification, achieving a higher SNR than conventional metal antennas. D. Arumugam \textit{et al.} \cite{arumugam2025stark} employed room-temperature Rydberg dissipative time-crystal (DTC) limit-cycle oscillations modulated by DC/AC Stark fields, enabling high-sensitivity detection in the DC-600 Hz sub-kHz regime. D. Xiao \textit{et al.} \cite{xiao2025optical} constructed a Cs ladder-type two-photon three-level system and realized accurate ULF electric-field measurement through a matched-measurement scheme and electric-field inversion. M. Chen \textit{et al.} \cite{chen2026calibration} further established a combined DC and AC-Stark response model, enabling calibration-free electric-field measurement in the sub-MHz regime. To improve VHF-band detection, N. Prajapati \textit{et al.} \cite{prajapati2024high1} proposed the HAMMER method, which utilizes high-angular-momentum Rydberg-state coupling to enhance low-frequency sensing performance.

In practical low-frequency operation, vapor-cell screening is a critical challenge. To address this, Y. Y. Jau \textit{et al.} \cite{jau2020vapor} adopted a sapphire Rb vapor cell with high sheet resistance, extending the electric-field screening time and enabling linear-response measurements of sub-kHz fields. Y. Xie \textit{et al.} \cite{xie2025low} designed an electrode-embedded Cs vapor cell and demonstrated multi-modulation communication with a 100 kHz carrier based on EIT and the AC Stark effect. M. Lei \textit{et al.} \cite{lei2024high} achieved high-sensitivity detection in the ULF, VLF, and LF bands by optimizing parallel electrode plates and DC bias conditions. Beyond field measurement, some studies focus on low-frequency communication. Shi \textit{et al.} \cite{shi2025new} utilized multiple harmonic signals for phase detection of low-frequency signal in the 30-300 kHz band, and demonstrated its image transmission capability. Zhang \textit{et al.} \cite{zhang2026self} built a self-dressed superheterodyne inside the sapphire vapor cell based on a undesired laser-induced DC field, enabling high-sensitivity kHz-band signal reception. Qiu \textit{et al.} \cite{qiu2026shortwave} further demonstrated shortwave frequency-hopping reception using a RAQ-based electrically small antenna, realizing communication across the 3-30 MHz HF band.

The performance comparison of typical low-frequency signal measurement schemes is summarized in Fig.~\ref{fig:Frequency range}.

\section{Equivalent Channel Model and Performance Analysis of RAQ-Based Communications}

Based on the physical fundamentals and system architecture of RAQ receivers, this section reviews and analyzes the research on equivalent channel models and the theoretical performance of RAQ communication systems. The discussion is divided into four aspects: the physical mapping from modulated  signal to atomic response, the modeling of equivalent communication channels, the theoretical analysis of communication capacity, and the non-ideal sources, which connect quantum response mechanisms with communication theory and provide a foundation for further exploration of RAQ communication technologies.

%\subsection{Degrees of Freedom}
%\subsection{Parametric Mapping From modulated signal to Atomic Response}
\subsection{Mapping From Modulated Signal to Atomic Response}

Based on the atomic response and readout methodology in Section \ref{sec:RAQ Transceiver System}, this section surveys how the modulated information of RF signal map onto the atomic response and converted into detectable features of the output probe laser, which bridges equivalent channel models with the physical implementation of information carrying in atomic communication links.

%\subsubsection{Characteristic of the EIT Spectrum for Information Carrying} 
\subsubsection{Typical Features of Probe Transmission Spectrum}
%\subsubsection{Observable Feature Space of the Optical Readout}

For the RAQ receiver, the transmission spectrum obtained by scanning frequencies around the probe resonance serves as the optical output carrier for the RF modulated information, which determine the specific form of the received signal. RF-atom interactions perturb this transmission spectrum across multiple dimensions, and the recovery of diverse modulated information depends on extracting features from distinct spectral dimensions. Overall, the features of transmission spectrum used for signal detection span the horizontal axis, i.e., frequency, and the vertical axis, i.e., transmittance.
﻿

\begin{figure*}[!t]
	\centering
	\includegraphics[width=0.8\textwidth]{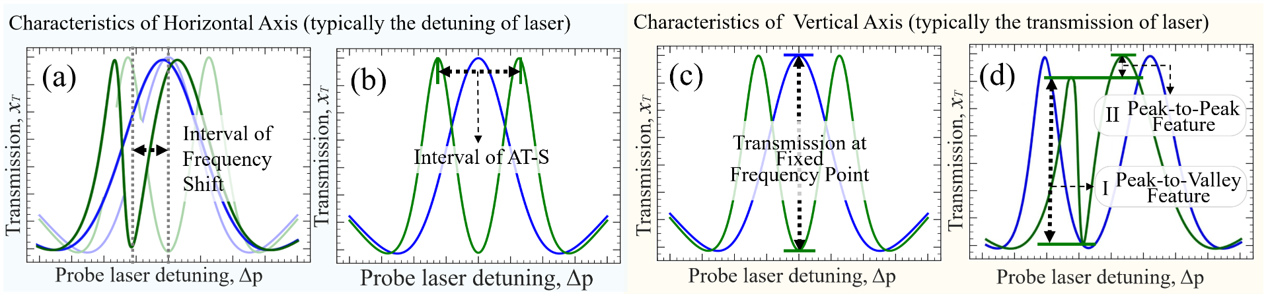}  
	\caption{Characteristic of the EIT spectrum for information carrying. (a) The interval of frequency shift. (b) The interval of Autler-Townes splitting. (c) The transmittance at a fixed laser frequency. (d) The transmission of the two asymmetric AT splitting peaks and the transmission of difference between a peak and its adjacent valley. Blue lines: transmission spectrum without RF signal. Green lines: transmission spectrum with RF signal. }
	\label{fig_feature}
\end{figure*}

From the horizontal axis of the transmission spectrum, the observable characteristics primarily include two aspects: 
\begin{itemize}
	\item The interval of frequency shift. It refers to the displacement of the EIT transparency window center or the split  doublet center on the frequency axis \cite{jing2020atomic,schlossberger2024rydberg,gordon2014millimeter}, as illustrated in Fig. \ref{fig_feature}a, with the magnitude of this displacement positively correlated with the frequency of the RF field.
	\item The interval of AT splitting. It refers to the frequency separation between the two split peaks when the transmission spectrum undergoes splitting under the RF field, as illustrated in Fig. \ref{fig_feature}b. This interval is positively correlated with the amplitude of the RF field.
\end{itemize}
By detecting the changes in these two characteristic parameters of the transmission spectrum, it is possible to recover the frequency and amplitude of the RF field.

Information can also be acquired from the vertical axis of the transmission spectrum, including the following two aspects.
\begin{itemize}
	\item The transmittance at a fixed probe laser frequency \cite{simons2019rydberg,chen2025harnessing}. Amplitude or frequency variations of the RF field cause shifts of the EIT or the AT splitting window, thereby altering the transmission, as illustrated in Fig. \ref{fig_feature}c. 
	\item The difference of transmittance under asymmetric AT splitting. When signal detuning is introduced, the transmission spectrum deforms and becomes asymmetric. In such cases, information can be extracted by detecting the transmission of the two asymmetric AT splitting peaks, i.e., the peak-to-peak feature, as indicated in Region~$\mathrm{I}$ of Fig. \ref{fig_feature}d. Simultaneously, the information carried by the signal can also be obtained by measuring the transmission difference between a peak and its adjacent valley, i.e., the peak-to-valley feature, shown in Region~$\mathrm{II}$ of Fig. \ref{fig_feature}d. 
\end{itemize}

In summary, deeper investigation into the available features of the transmission spectrum can provide innovative insights and approaches for enhancing the capability of RAQ receivers.

%\subsubsection{Rydberg Reception of Different Information Modulation Schemes}
%\subsubsection{Projection of the Signal Modulation Space}
\subsubsection{Atomic Response Features-Based Modulation Mapping}
%{\color{red} In current research on RAQ receivers, information is typically modulated in dimensions such as amplitude, frequency, phase, and polarization based on the direct or indirect utilization of transmission spectrum characteristics.}
Information modulation in RAQ receivers is generally realized by mapping RF signal parameters onto measurable atomic-response features in the optical readout. Depending on the architecture and readout methodology, modulated information including amplitude, frequency, phase, and polarization can be encoded into AT peak separation, transmission variation, spectral asymmetry, or optical phase shift. %Thus, RAQ receivers convert RF-induced atomic-state perturbations into detectable optical observables rather than directly measuring the RF waveform as conventional antennas do.

The amplitude of the RF signal can be extracted from either the AT splitting interval or the probe laser transmission variation, as shown in Fig. \ref{fig_feature}b and c. In the resolvable AT-splitting regime, the separation between the two EIT-AT transmission peaks is proportional to the incident RF electric-field amplitude. In the unresolved or weak-field regime, the variation of the transmission is also approximately linear with the RF field strength around a properly selected operating point \cite{anderson2021atomic, song2018quantum}. Although both methods originate from RF-induced perturbations of the EIT spectrum, they differ in detection complexity, sensitivity, dynamic range, and response speed.

RF frequency information can be recovered from frequency-dependent changes in the transmission spectrum, including spectral shifts, peak asymmetry, and contrast variations \cite{kumar2017rydberg,anderson2017continuous}. As illustrated in Fig.~\ref{fig_feature}a, when the RF frequency matches a neighboring Rydberg transition, the RF field produces a EIT-AT response. When the RF frequency is detuned from the transition resonance, the EIT-AT peaks shift and become asymmetric due to the detuning-dependent dressed-state response. Therefore, frequency variations can be demodulated by scanning the laser detuning and tracking the spectral features, or by fixing the laser detuning at a sensitive operating point and converting the frequency-dependent spectral response into probe-transmission variation under a constant RF amplitude. In addition, the contrast and relative height of the AT-splitting peaks can also serve as frequency-dependent observables.

The measurement of RF phase information generally requires a phase reference. Existing approaches can be broadly divided into two categories. The first approach introduces an additional LO with a known phase that mixes with the RF signal through the atomic medium, converting the phase information into the amplitude of their beat signal. This beat signal can then be read out using conventional amplitude detection methods\cite{cai2023high, jing2020atomic, simons2019embedding}. The second approach uses optical-domain references, for example by modulating the coupling laser phase to generate optical sidebands that act as intrinsic reference fields without requiring an external RF LO \cite{anderson2022optical,anderson2020rydberg}. Both methods Both approaches exploit the atomic medium as a coherent mixer, converting RF phase information into detectable optical amplitude or phase variations.

Polarization information can be recovered by mapping the vector properties of the RF field onto polarization-dependent optical responses. Existing studies mainly follow two approaches. The first introduces a set of LO fields with specific polarization directions so that the incident RF field can be projected onto these reference axes and reconstructed from the corresponding transmission-spectrum features \cite{elgee2024complete}. The second exploits the dependence of  energy levels and transition strengths on the RF field polarization, such as Stark or Zeeman mechanisms. In this case, polarization information can be inferred directly from spectral shifts, splitting patterns, or transition strengths without reference fields \cite{wang2023precise, chen2025polarization}. The essence of both approaches is to transform polarization information, which is difficult to capture directly, into quantifiable optical observables.

\subsection{Atomic-to-Baseband Equivalent Channel Modeling}

The atomic response model discussed in the previous section describes the mapping from the incident RF field to the atomic coherence, optical susceptibility, and probe transmission. For wireless communication analysis, however, it is necessary to further translate this atomic response into an equivalent input-output relationship. In this view, the RAQ receiver is treated as communication front end that converts RF waveform into an electrical baseband signal through atomic transduction, optical readout, photodetection, and baseband processing.

Denote $x(t)$ as the complex baseband envelope, with the corresponding incident RF field $E_s(t)=\text{Re}\{x(t)e^{j2\pi f_ct}\}$ and Rabi frequency $\Omega_s(t)=\mu_sE_s(t)/\hbar$. 
The atomic-to-baseband equivalent channel can be generally written as
%The atomic-to-baseband equivalent channel can be generally written as
\begin{equation}
	y(t)=\mathcal{H}_{\rm eq}\left(x(t)\right)+n_{\rm eq}(t), \label{eq:general_io_model}
\end{equation}
where $y(t)$ is the received baseband output that depends on corresponding optical or electrical readout. $\mathcal{H}_{\rm eq}(\cdot)$ denotes the equivalent channel response that absorbs both the atomic response, optical readout and electrical detection chain. The term $n_{\rm eq}(t)$ denotes the noise sources according to \ref{subsubsec:noise model}.

\begin{figure*}[!t]
	\centering
	\includegraphics[width=\textwidth]{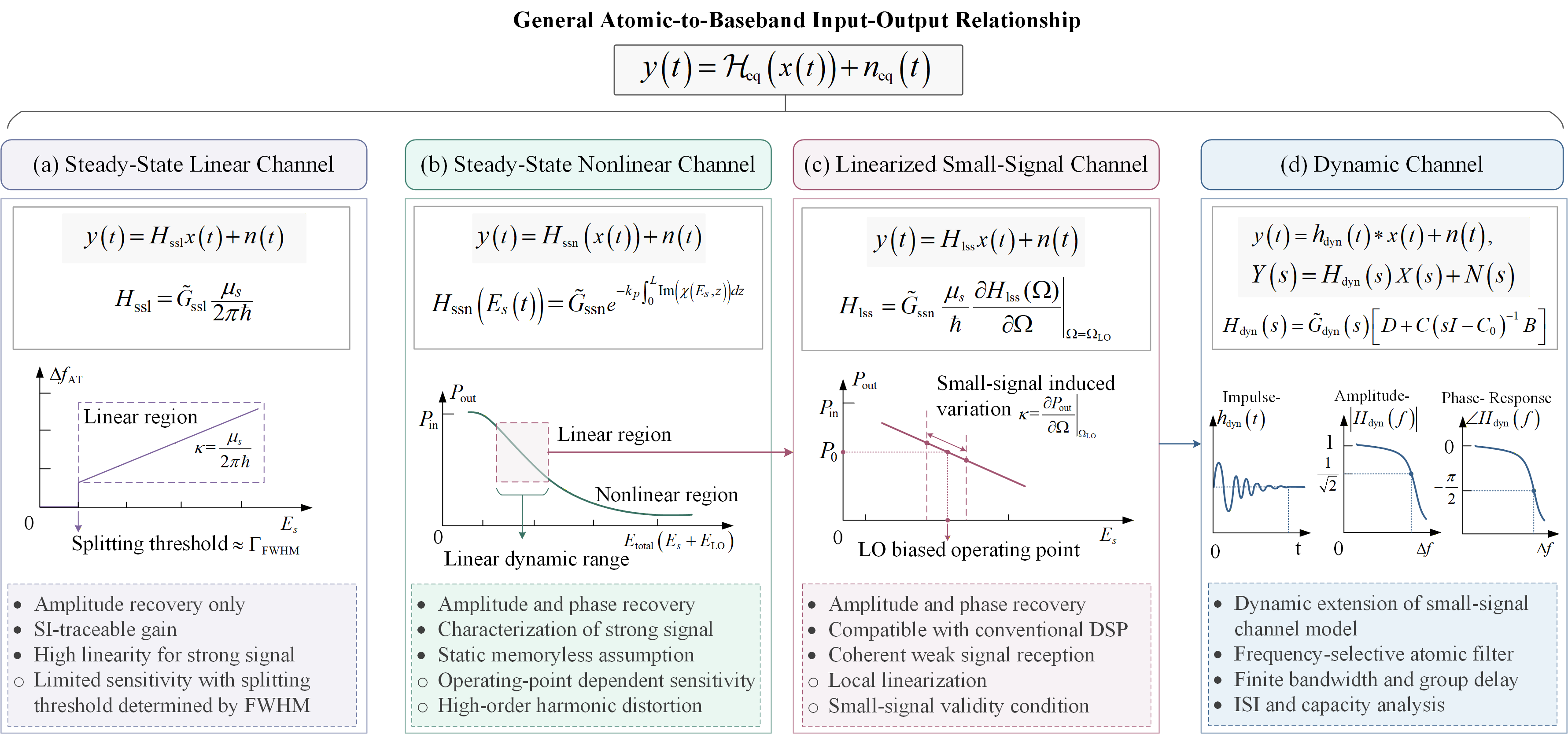}
	\caption{Atomic-to-Baseband Equivalent Channel Modeling. (a) Steady-State Linear Channel \cite{jing2020atomic,schlossberger2024rydberg,holloway2017electric,anderson2021self}. (b) Steady-State Nonlinear Channel \cite{simons2019rydberg,chen2025harnessing}. (c) Linearized Small-Signal Model \cite{gong2026rydberg_models,prajapati2022tv}. (d) Dynamic Model \cite{bohaichuk2022origins,zhu2025general,han2026rydberg}.}
	\label{Fig:Equivalent Channel Modeling}
\end{figure*}

Existing studies have explored equivalent channel models for different operating scenarios and conditions, mainly including the steady-state linear, steady-state nonlinear, linearized small-signal, and dynamic channels, as shown in Fig.~\ref{Fig:Equivalent Channel Modeling}.

\subsubsection{Steady-State Linear Channel}

When the RAQ receiver is driven by a strong RF signal, the system operates in the resolvable AT-splitting regime \cite{anderson2016optical}. In this case, the received communication observable is not a directly sampled baseband waveform, but a spectral feature extracted from the EIT spectrum. If the RF envelope slowly varying, the output can be represented by the spacing of AT splitting $y(t)=\Delta f_{\rm AT}(t)$ \cite{jing2020atomic,schlossberger2024rydberg}.  Since AT splitting measures the RF-field amplitude rather than the complex envelope phase, the atomic-to-baseband equivalent channel is linear with respect to the amplitude envelope, i.e.,
\begin{equation}
		y(t) = H_{\rm ssl}|E_s(t)|+n(t), \quad H_{\rm ssl}=\tilde{G}_{\rm  ssl}\frac{\mu_s}{2\pi\hbar}.
\end{equation}
Here, $\mu_s/2\pi\hbar$ is the physical field-to-splitting coefficient, while $\tilde{G}_{\rm ssl}$ is the splitting-to-baseband conversion gain determined by the spectral readout, frequency estimation, calibration, and subsequent electronic or digital processing.
%The optical collection efficiency, photodetector responsivity, transimpedance gain, and amplifier gain mainly affect the variance of the splitting-estimation error, rather than the physical field-to-splitting coefficient itself.

This model exhibits a direct amplitude-to-frequency mapping between incident RF field and the measured spectral splitting. The channel response $H_{\rm ssl}$ depends only on Planck's constant $\hbar$ and the atomic dipole moment $\mu_s$. Therefore, it offers the key advantages of SI-traceable and highly-linearity   \cite{holloway2017electric,anderson2021self}.  However,  this model is mainly applicable to slowly varying AM-envelope detection or strong-field, and the AT splitting must be larger than the full width at half maximum (FWHM) of EIT window $\Gamma_{\rm FWHM}$ to be resolved, i.e., $\Omega_{s} \gtrsim \Gamma_{\rm FWHM}$.

\subsubsection{Steady-State Nonlinear Channel}
When the RF field is weak, i.e., $\Omega_{s} \ll \Gamma_{\rm FWHM}$, the AT splitting remains unresolved. 
In this case, the received signal is usually obtained from the transmitted probe intensity $P_{\rm out}$ or phase  $\Delta\phi$ at a fixed operating point. Under the steady-state approximation, the RAQ receiver can be modeled as a nonlinear memoryless channel,
\begin{equation}
		y(t)=H_{\rm ssn}(x(t))+n(t),
		\label{eq:steady-state nonlinear}
\end{equation}
where $H_{\rm ssn}(\cdot)$ is determined by the steady-state solution of the OBEs, the readout chain. For the optical intensity readout, the steady-state nonlinear mapping from RF field to probe transmission can be expressed as
%where $x$ denotes the RF field amplitude $E_{s}$ or Rabi frequency $\Omega_{s}$, and $y$ can be the transmitted probe power $P_{\rm out}$ or the phase shift $\Delta\phi$, depending on the readout method. The equivalent nonlinear channel of RAQ receiver is given by \cite{simons2019rydberg,chen2025harnessing}
\begin{equation}
		H_{\rm ssn}(E_s(t))
		=
		\tilde{G}_{\rm ssn}e^{
		-k_p\int_{0}^{L}
		\operatorname{Im}
		\left(
		\frac{2N_0|\mu_{eg}|^2}{\varepsilon_0\hbar\Omega_p(z)}
		\rho_{21}(E_s(t),z)
		\right) dz
	}.
\end{equation}
Here, $\tilde{G}_{\rm ssn}=\eta\alpha_{\rm pd}G_{\rm TIA}G_{\rm LNA}G_{\rm bb}$ denotes the transmission-to-baseband conversion gain, including the optical collection efficiency $\eta$, photodetector responsivity $\alpha_{\rm pd}$, transimpedance gain $G_{\rm TIA}$, low-noise/baseband amplifier gain $G_{\rm LNA}$, and the baseband processing gain $G_{\rm bb}$.
For phase readout, the output can be similarly constructed from $\text{Re}\left(\chi\right)$ with the corresponding phase-to-baseband conversion gain.

This model captures the static nonlinear amplitude-to-intensity or amplitude-to-phase conversion of the atomic medium. It is useful for analyzing operating-point selection, dynamic range, saturation, nonlinear distortion, and 1-dB compression. However, without an external phase reference, this channel mainly supports envelope-type detection rather than coherent complex-baseband reception.

\subsubsection{Linearized Small-Signal Channel}
For a LO-dressed or superheterodyne reception, the nonlinear steady-state response can be locally linearized around an LO-biased operating point \cite{gong2026rydberg_models,prajapati2022tv}. Let the total incident field be composed of a strong LO and a weak communication signal,
$E_{\rm total}(t)	=
	\text{Re}
		\{E_{\rm LO}+x(t)e^{j2\pi f_{\rm IF}t} \}e^{j2\pi f_{\rm LO}t}$,
where $E_{\rm LO}$ sets the operating point. The Rabi frequency is
$\Omega_{\rm total}(t)
	=
	\Omega_{\rm LO}
	+
	\delta\Omega_s(t),
	|\delta\Omega_s(t)|\ll |\Omega_{\rm LO}|.$
Applying a first-order Taylor expansion to the steady-state nonlinear mapping around the LO operating point gives
\begin{equation}
	y(t)
	\approx
	H_{\rm lss}(\Omega_{\rm LO})
	+
	\left.
	\frac{\partial H_{\rm lss}(\Omega)}
	{\partial \Omega}
	\right|_{\Omega=\Omega_{\rm LO}}
	\delta\Omega_s(t)
	+
	n(t).
\end{equation}
After removing the DC component and downconverting the IF beat note, the baseband input--output relationship becomes
\begin{equation}
	y(t)
	=
	H_{\rm lss}x(t)
	+
	n_{\rm eq}(t),
	\label{eq:lss_bb_channel}
\end{equation}
where the local small-signal gain is
\begin{equation}
	H_{\rm lss}
	=
	\left.
	\tilde{G}_{\rm ssn}\frac{\partial H_{\rm lss}(\Omega)}
	{\partial \Omega}
	\right|_{\Omega=\Omega_{\rm LO}}
	\frac{\mu_s}{\hbar}.
	\label{eq:lss_gain}
\end{equation}
%Here, $G_{\rm ro}$ denotes the equivalent gain of the optical readout, photodetection, and electronic processing chain. 
This model converts the nonlinear atomic response into a local linear channel, making it closest to the conventional baseband channel model in wireless communications, and enabling synchronization, channel estimation, equalization, demodulation, and DSP processing. However, it is valid only within the small-signal region around the selected operating point; otherwise, higher-order nonlinear terms lead to gain compression and distortion.

\subsubsection{Dynamic Channel}
The preceding models are static channel models and are valid only when the signal bandwidth is much smaller than the atomic relaxation rates. For high-speed communication signals, the time-dependent atomic response must be considered, and a dynamic channel model is required. 

Under a fixed operating point and the small-signal approximation, the nonlinear RAQ receiver can be locally approximated as a linear time-invariant (LTI) system. By linearizing the time-dependent OBEs around the steady-state solution, the perturbed atomic dynamics can be written in a state-space form as $
\delta\dot{\mathbf{z}}(t)
=
C_0\delta\mathbf{z}(t)
+
Bx(t)$, $
\delta y(t)
=
C\delta\mathbf{z}(t)
+
D x(t)
+
n(t)$, 
where $\delta\mathbf{z}(t)$ denotes the perturbation of the atomic state vector, matrix $C_0$ is determined by the linearized Liouvillian dynamics, and $B, C, D$ are coefficient matrices based on system parameters \cite{zhu2025general,han2026rydberg}.
Taking the Laplace transform gives the dynamic input-output relationship
\begin{equation}
	Y(s)
	=
	H_{\rm dyn}(s)X(s)
	+
	N(s),
\end{equation}
where the equivalent dynamic channel response is
\begin{equation}
	H_{\rm dyn}(s)
	= \tilde{G}_{\rm dyn}(s)
	\left[
	D
	+
	C(sI-C_0)^{-1}B\right].
\end{equation}
Here, $\tilde{G}_{\rm dyn}(s)=\eta\alpha_{\rm pd}G_{\rm TIA}G_{\rm LNA}G_{\rm bb}(s)$ is the frequency response, which shows that the overall baseband bandwidth is jointly limited by the atomic dynamics and the electrical readout chain.
Equivalently, in the time domain, the dynamic channel can be expressed as
$
y(t)
=
h_{\rm dyn}(t)*x(t)
+
n(t)$,
where $h_{\rm dyn}(t)$ is the impulse response corresponding to $H_{\rm dyn}(s)$ \cite{bohaichuk2022origins}.

The dynamic channel response generally exhibits low-pass characteristics, with the finite lifetimes of Rydberg states, EIT settling time, and decoherence processes jointly determined the receiver bandwidth, amplitude attenuation, and phase delay. 
%Recent studies have introduced quantum transconductance to characterize this conversion efficiency \cite{zhu2025general}. 
It enables quantitative analysis of high-speed reception, inter-symbol interference, group delay, and capacity limits of RAQ receivers.

\subsection{Communication Capacity}

As channel capacity serves as one of the most critical metrics for evaluating system performance in communications, it has emerged as a subject of increasing research interest within the paradigm of RAQ receivers.

Early research focused on establishing fundamental capacity scaling laws. For an array of RAQ receivers operating in a single-input-multi-output (SIMO) scenario, the data capacity was shown to scale as $B\log_2(1+N\times\mathrm{SNR})$ for $N$ receivers \cite{otto2021data}. This result provides a basic understanding of capacity enhancement via spatial diversity in distributed RAQ receivers.

The analysis was later extended to multi-antenna systems. 
Yuan \textit{et al.} extracted antenna properties from the quantum characteristics of atoms and incorporated them into electromagnetic modeling for RAQ MIMO systems \cite{yuan2025electromagnetic}. This work enables MIMO capacity analysis under spatial multiplexing scenarios and shows that RAQ receiver arrays can provide certain capacity advantages over classical dipole arrays under the same SNR condition.

Further studies investigated the information-theoretic constraints of RAQ reception. The capacity dilemma of magnitude-only RAQ receivers was analyzed by deriving the mutual information between the transmitted signal and the magnitude-only received signal \cite{guo2025high}. The results show that although the loss of phase information reduces mutual information, the thermal-noise immunity of RAQ receivers can still provide a capacity advantage over traditional RF receivers, especially when reference signals are employed.

In addition to studies that directly calculate communication capacity, some works evaluate communication performance through through SNR or dynamic-response analysis. Gong \textit{et al.} developed an equivalent baseband signal model for RAQ receivers and quantified their SNR advantages over classical RF receivers \cite{gong2026rydberg_models}. Zhu and Dai proposed a general dynamic signal model and introduced quantum transconductance to characterize the dynamic response and in-band noise effects of RAQ receivers \cite{zhu2025general}. Although these studies do not directly derive capacity expressions, their analysis provide a basis for capacity evaluation under different operating conditions.

Note that for RAQ-based communication systems, capacity is not only determined by sensitivity or received SNR, but also strongly constrained by the bandwidth, especially the limited instantaneous bandwidth of atomic response.

\subsection{Non-Ideal Sources in RAQ Receiver}

In this subsection, we analyze major impairments and interference in practical RAQ reception, together with representative mitigation methods proposed in the literature.

\subsubsection{Channel Impairments} Channel impairments in RAQ receivers mainly originate from the nonideal RF-to-optical transduction of the atomic medium. Typical effects include the intrinsic nonlinear EIT transmittance, higher-order mixing between signal and LO fields, multi-field Rabi-frequency superposition, Doppler broadening induced by thermal atomic motion, Floquet-induced harmonic generation, and detuning-induced spectral asymmetry \cite{xia2024digital,wu2023linear,wu2025analysis,zhu2025raq,goncalves2024nonlinearities,zou2020atomic,legaie2024millimeter,kasza2025atomic}..

\begin{itemize}
	\item \textit{Inherent Nonlinear Response}: The response of RAQ receivers becomes nonlinear in the strong-signal region, leading to signal distortion and performance degradation if linear assumptions are still applied in system design \cite{xia2024digital}. To mitigate this effect, Xia \textit{et al.} \cite{xia2024digital} develop a nonlinear compensation framework based on a quartic function model. Wu \textit{et al.} \cite{wu2023linear} start from the expression of density matrix elements to evaluate the operating conditions for maximizing the linear dynamic range. Alternatively, Wu \textit{et al.} \cite{wu2025analysis} employ the Bussgang theorem to characterize signal nonlinear distortion.
	
	\item \textit{Superposition of Rabi Frequencies}: The superposition of Rabi frequencies can introduce high-order nonlinear terms arising from multi-photon interactions in RAQ receivers, particularly under multi-field excitation. To address this, Gonçalves \textit{et al.} \cite{goncalves2024nonlinearities} analyze harmonic and intermodulation distortion, and develop a nonlinear model based on master equations to characterize and suppress such high-order distortions.
	
	\item \textit{Doppler Broadening}: Doppler broadening arises from the thermal motion of atoms, which causes frequency shifts due to the velocity-dependent detuning of atomic transitions. This effect distorts the EIT-AT spectrum and introduces additional nonlinearity in RAQ receivers.
	Gonçalves \textit{et al.} \cite{goncalves2024nonlinearities} perform numerical simulations to analyze the impact of Doppler broadening on the power spectrum of harmonic and intermodulation products, indicating that careful selection of physical parameters can mitigate the resulting nonlinearities.
	
	\item \textit{Floquet-Induced Nonlinearity}: A large power disparity between the LO and signal field can drive the atomic system into a Floquet regime, generating high-order harmonics through nonlinear wave mixing \cite{miller2016radio,anderson2014twophoton}. These harmonics increase the noise floor and introduce signal distortion, imposing a trade-off between instantaneous bandwidth and sensitivity. To overcome this limitation, Qimeng \textit{et al.} \cite{Qimeng2026bw} propose a dual-wave mixing matching optimization that jointly tunes the Rabi frequencies of both LO and signal fields.

	\item \textit{Detuning}: Detuning occurs when signal or laser frequencies deviate from the atomic resonance, causing EIT-AT spectral deformation and introducing additional distortion. Zou \textit{et al.} \cite{zou2020atomic} mention nonlinear distortion introduced by spectral asymmetry due to detuning, while Legaie \textit{et al.} \cite{legaie2024millimeter} investigate the impact of coupling laser detuning and LO detuning on RAQ receiver performance.
\end{itemize}

\subsubsection{Interference} RAQ receivers also suffer from several types of interference, including blocking from strong external interference, inter-carrier interference (ICI), and intermediate-frequency interference (IFI) \cite{wu2023linear,zou2020atomic,zhu2025raq}.

\begin{itemize}
	\item \textit{Blocking}: Blocking occurs when strong interference saturates the atomic response, whcih drives the receiver out of its linear operating region, suppresses the target signal and potentially renders it undetectable. Wu \textit{et al.} power \cite{wu2023linear} quantify this effect by introducing a single-tone interference signal and evaluating its impact on the effective intermediate-frequency signal.
	
	\item \textit{Inter-Carrier Interference (ICI)}: ICI results from interactions between different carrier frequencies, degrading multi-carrier communication performance. Zou \textit{et al.} \cite{zou2020atomic} demonstrate that appropriate carrier spacing can effectively mitigate this interference in atomic receivers.
	
	\item \textit{Intermediate-Frequency Interference (IFI)}: IFI occurs when signals from different frequency bands overlap at the same intermediate frequency, causing distortion. Zhu \textit{et al.} \cite{zhu2025raq} effectively mitigate this effect via a multi-cell atomic receiver architecture and qWMMSE detection.
\end{itemize}

Overall, channel impairments and interference constitute the major non-ideal sources in practical RAQ receivers. Their accurate characterization and mitigation are critical for improving the robustness, reliability, and performance of such systems under practical communication scenarios.

\section{RAQ Radio-Enhanced Advanced Communication Technologies}
The unique physical properties and performance advantages of RAQ radio empower diverse wireless communication technologies. This section begins with the implementation of advanced modulation schemes to introduce the feasibility of atomic information carriage. By utilizing the high-sensitivity of RAQ receiver, it is naturally extended to sensing and cognitive communications for enhanced spectral awareness. Meawhile, the intrinsic frequency selectivity of atomic transitions can be exploited to achieve interference-resilient communications in complex electromagnetic environments. Furthermore, bypassing classical hardware constraints endows RAQ receiver with full-spectrum agility across extreme high-frequency and low-frequency communications. Besides, the atomic front-ends can also be integrated into MIMO architectures to fully exploit spatial degrees of freedom.

\subsection{Implementation of Advanced Modulation Schemes}

\begin{table*}[t]
	\caption{Classification of Modulation Schemes for RAQ Receivers}
	\label{tab:feasibility_exploration}
	\centering
	\setlength{\tabcolsep}{6pt} % 稍微加宽列间距，视觉效果更舒展
	\begin{tabular}{llll} % 顶刊标准：去掉所有竖线 "|"
		\toprule
		\textbf{Category} & \textbf{Modulation Attribute} & \textbf{Specific Scheme} & \textbf{References} \\
		\midrule
		\multirow{3}{*}{Non-coherent} & Amplitude & AM & \cite{anderson2020rydberg, anderson2021atomic, song2018quantum, yuan2024rydberg, anderson2024long} \\
		\cmidrule{2-4}
		& \multirow{2}{*}{Frequency} & FM & \cite{anderson2021atomic, schlossberger2025rydberg, kumar2017rydberg, anderson2024long} \\
		&                            & 4FSK & \cite{gao2025rydberg} \\
		\midrule
		\multirow{4}{*}{Coherent}     & \multirow{3}{*}{Phase}     & MPSK & \cite{anderson2020rydberg, simons2019rydberg, simons2019embedding, prajapati2023synthetic, jing2020atomic, cai2023high, Wang2026apsk} \\
		&                            & BPSK & \cite{holloway2019detecting, anderson2024long} \\
		&                            & QPSK & \cite{holloway2019detecting} \\
		\cmidrule{2-4}
		& Hybrid           & QAM  & \cite{holloway2019detecting, nowosielski2024warm} \\
		\midrule
		\multirow{2}{*}{Others} & Polarization               & Polarization modulation   & \cite{wang2023precise, elgee2024complete, cloutman2024polarization, ding2024circularly} \\
		\cmidrule{2-4}
		& Orbital angular momentum                        & OAM-FDM, OAM & \cite{Wang2026oamfdm}\cite{wang2026quadrupole} \\
		\bottomrule
	\end{tabular}
\end{table*}

Early explorations of RAQ receivers focused on amplitude and frequency detection for non-coherent systems \cite{sedlacek2012microwave}, laying the foundation for RAQ-enhanced communication technologies. In recent years, advances in phase detection have extended RAQ receivers towards coherent systems. Moreover, polarization detection further provides a new dimension to improve the information capacity of RAQ-enhanced systems.

RAQ receivers exhibit direct response to electromagnetic-field amplitude and frequency, therefore naturally support AM and FM reception. For AM, Anderson \textit{et al.} \cite{anderson2020rydberg, anderson2021atomic} verified the feasibility of Rydberg atoms for AM radio communication. Song \textit{et al.}\cite{song2018quantum} extended this work and demonstrated a modulation bandwidth of 1 MHz and a data rate of approximately 1 Mbps. To improve AM reception capability, Yuan \textit{et al.}\cite{yuan2024rydberg} adopted a dual-tone microwave scheme for increased capacity and bandwidth. For FM,  Anderson \textit{et al.}\cite{anderson2021atomic} demonstrated RAQ-based FM reception and Schlossberger \textit{et al.} \cite{schlossberger2025rydberg} further promoted this by receiving real FM audio signals from commercial handheld radios. Kumar \textit{et al.}\cite{kumar2017rydberg} employed FM spectroscopy to optimize RAQ-based electric field measurement. For more complex  electromagnetic environments , Gao \textit{et al.} \cite{gao2025rydberg} implement 4-state frequency-shift keying digital communication with time-varying radar interference.

The development of phase detection further expanded RAQ receivers from non-coherent to coherent communication systems. Anderson \textit{et al.} \cite{anderson2020rydberg}  developed phase detection method for pulsed RF fields, including a scheme that does not require an external RF reference. Simons \textit{et al.} \cite{simons2019rydberg} proposed a Rydberg-atom mixer to realize signal down-conversion, and the same team integrated Rydberg sensors into antennas for simultaneous amplitude and phase detection \cite{simons2019embedding}. For a 28 GHz far field application, Prajapati \textit{et al.} \cite{prajapati2023synthetic} utilized synthetic aperture techniques to achieve high-resolution phase detection. Based on phase reception, RAQ systems have been extended to hybrid and higher-order modulation schemes, including PSK and QAM. Holloway \textit{et al.} \cite{holloway2019detecting} demonstrated RAQ-based reception for BPSK, QPSK and QAM signals. Jing \textit{et al.} \cite{jing2020atomic} presented the atomic superheterodyne receiver for phase and frequency measurement with enhanced detection sensitivity, and Cai \textit{et al.} \cite{cai2023high} realized high-sensitivity reception matched with frequency division multiplexing (FDM). Furthermore, Nowosielski \textit{et al.}  \cite{nowosielski2024warm} developed a warm Rydberg-atom QAM receiver supporting a channel capacity of 19.3 Mbps. Anderson \textit{et al.}  \cite{anderson2024long} built quantum radios for long-distance transmission to receive AM, FM and BPSK signals. To expand modulation compatibility, Wang \textit{et al.} \cite{Wang2026apsk} leveraged the self-demodulation feature of Rydberg atoms to decode APSK microwave signals, and analyzed the relationship between modulation orders and BER, symbol duration, and image-transmission qualityn.

Other information-carrying dimensions include polarization and orbital angular momentum (OAM) \cite{wang2026quadrupole} are also exploited in RAQ receivers. For polarization reception, Wang \textit{et al.} \cite{wang2023precise} demonstrated precise microwave polarization measurement using Rydberg atom-based mixers, while Elgee \textit{et al.} \cite{elgee2024complete} achieved three-dimensional vector polarimetry with Rydberg-atom electrometers. For circularly polarized signals, Ding \textit{et al.} \cite{ding2024circularly} proposed an omnidirectional Rydberg atomic sensor using characteristic mode analysis. Conversely, Cloutman \textit{et al.} \cite{cloutman2024polarization} investigated polarization-insensitive electrometry via specific Rydberg transitions, providing an alternative route for robust field detection under uncertain polarization states. For OAM-based communication, Wang and Zhang \cite{Wang2026oamfdm} proposed a digital communication framework that integrates OAM with FDM. By mapping each FDM subcarrier to a distinct OAM mode, they constructed a spectral–spatial two-dimensional multiplexing space, where OAM features are embedded into EIT spectra through quadrupole-gradient coupling, enabling simultaneous OAM identification and FDM demodulation.

\subsection{Sensing and Cognitive Communications}

\begin{figure*}[!t]
	\centering
	\includegraphics[width=\textwidth]{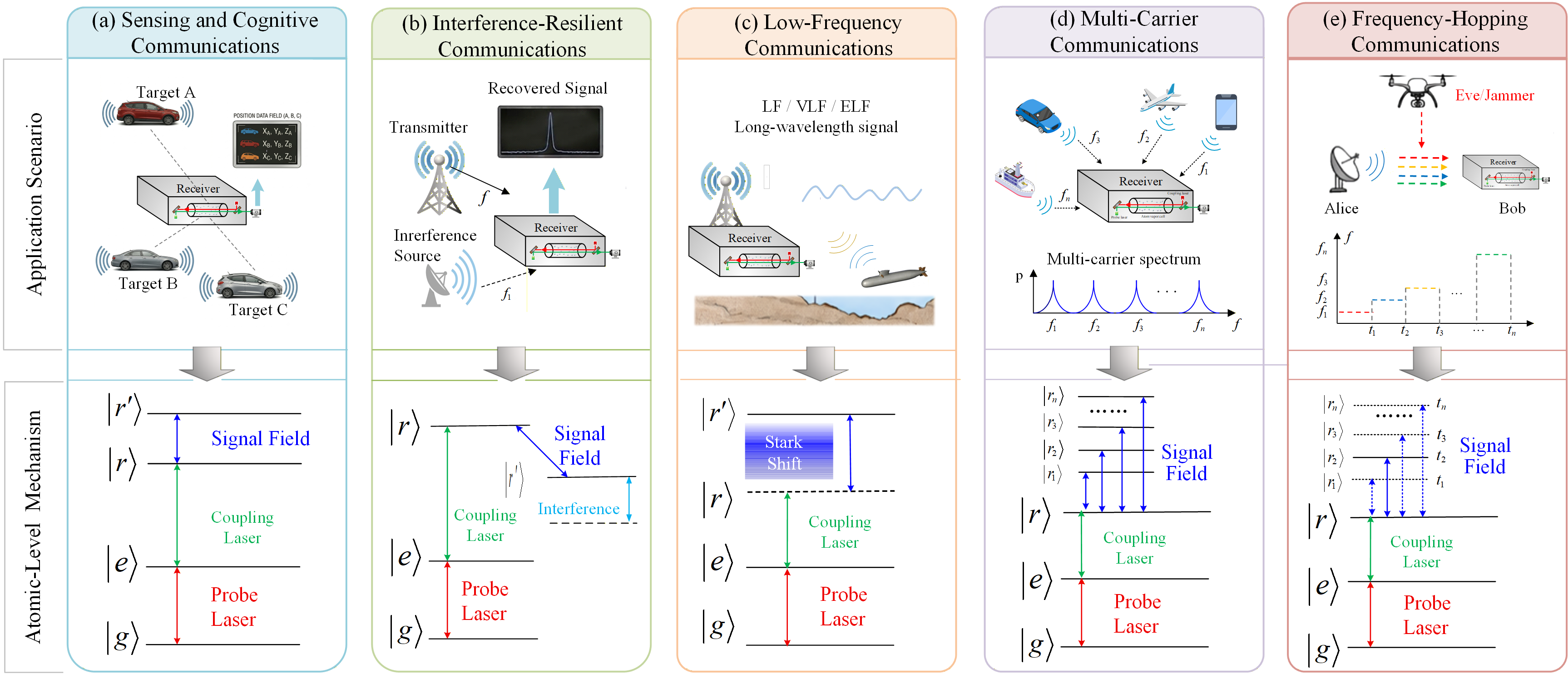}  % 设置为页面宽度的80%
	\caption{Advanced communication technologies enhanced by atomic radio. (a) Sensing and cognitive communications. (b) Interference-resilient communications. (c) Low-frequency communications. (d) Multi-carriers communications. (e) Frequency-hopping communications.}
	\label{fig: advanced technologies}
\end{figure*}

Owing to its inherent high-sensitivity to electromagnetic field, RAQ receiver has been applied to field sensing with better potential performance than traditional sensors \cite{holloway2017atom,holloway2014subwavelength,cardman2020atomic,anderson2018highresolution,anderson2023syntheticapertures}. Therefore, researchers are developing novel receiver architectures and algorithms to fully exploit the unique properties of atomic systems. Besides, advanced signal processing techniques are integrated to suppress noise and ensure robust system operation. The general mechanism of RAQ-enabled sensing is illustrated in Fig.~\ref{fig: advanced technologies}a.

To exploit the field-sensitive response of Rydberg atoms for spatial parameter estimation, Robinson \textit{et al.} \cite{robinson2021determining} inferred the angle-of-arrival (AoA) of incident RF waves via phase difference detection at two points in an atomic vapor cell. Further, Yan \textit{et al.}\cite{yan2025measurement} realized the simultaneous measurement of Doppler frequency shift (DFS) and AoA in a single Rydberg atom system. By employing dual-coupled beams to suppress measurement errors, their scheme achieved a DFS error within $\pm0.5$ Hz and an AoA error below $0.5^{\circ}$, while the relative frequency zero point enabled DFS direction discrimination. For multi-target scenarios, Han \textit{et al.} \cite{Han2026doa} expanded DoA estimation performance through an imaging-based spectral estimation method using a single RAQ receiver. By linearizing atomic absorption with a strong LO, they converted multi-target DoA estimation into a Prony-based spectral estimation problem and removed the dependence on vapor-cell length, thereby improving broadband applicability.

To further exploit the unique capabilities of RAQ receivers and enhance system performance, innovative receiver architectures have recently emerged. Towards multi-user sensing, Kim \textit{et al.} \cite{kim2025quantum} proposed Quantum-MUSIC, a multiple signal classification algorithm designed for quantum wireless sensing. By retrieving the channel matrix and integrating the MUSIC framework, it enables multi-user wireless sensing from magnitude-only measurements and outperforms classical approaches. Cui \textit{et al.} \cite{cui2025realizing} addressed the external reference requirement in heterodyne sensing by introducing a self-heterodyne paradigm, where transmitter self-interference serves as an embedded reference and the RAQ receiver functions as an atomic autocorrelator. This method supports wider bandwidth and approaches the Cramér-Rao lower bound, achieving high sensitivity over approximately 100 MHz bandwidth.

Advanced signal processing techniques, particularly deep learning, have also been employed to mitigate noise, compensate for non-ideal factors, and improve system performance. Xie \textit{et al.} \cite{xie2025atomic} combined a Rydberg sensor with a convolutional neural network (CNN)-transformer, where the CNN captures local temporal features and the transformer extracts global correlations, attenuating noise from laser fluctuations, frequency drift, and ambient conditions. This architecture demonstrated high accuracy across multifrequency microwave recognition, broadband spectrum monitoring, and automatic modulation classification. Similarly, Liu \textit{et al.} \cite{liu2022deep} developed a deep learning-enhanced receiver for multifrequency microwave recognition, directly decoding FDM signals while avoiding explicit solutions of the complex Lindblad master equation.

\subsection{Interference-Resilient Communications}

%RAQ receivers offer superior interference tolerance compared to conventional receivers based on their unique atom-resonance response and can be further enhanced dedicated algorithms for operation in contested electromagnetic environment.
RAQ receivers offer a unique physics mechanism of atomic transition-selective response. As illustrated in Fig.~\ref{fig: advanced technologies}b, the off-resonant signals are naturally suppressed during atomic transduction, thus interference rejection can occur before electronic filtering and baseband processing. Combined with dedicated algorithms and receiver architectures, this feature enables robust operation in contested electromagnetic environments.

The inherent frequency selectivity of atomic transtion, which functions as a built-in high-Q filter, is particularly suitable for anti-jamming communications. Rostampoor and Adve \cite{rostampoor2025interference} demonstrated that a Rb-85 based receiver detects 8-Pulse Amplitude Modulation (8-PAM) signals under off-resonant interference without additional filters, effectively functioning as an integrated filter-demodulator with lower symbol error rates than conventional receivers. Their work on ultra-wideband frequency-hopping spread spectrum (FHSS) systems further verified this advantage over 100 kHz to 20 GHz range, providing a 51 dB improvement in narrowband interference tolerance over single-frequency systems.

%To extend applicability and enhance performance further, new algorithms and architectures have also been investigated. 
Beyond passive interference rejection enabled by atomic transition selectivity, recent studies have also explored robust reception under practical conditions, where detection loss, multi-signal coexistence, and rapidly varying spectra may weaken the ideal filtering advantage.
Kurzyna \textit{et al.} \cite{kurzyna2025microwave} introduced quantum error correction techniques based on dipolar interactions between Rydberg atoms, which improves Fisher information by a factor of 3.3 and protecting sensing data against detection-stage losses. For multi-user scenarios, Wu \textit{et al.} \cite{wu2025characterization} analyzed mutual interference when multiple signals couple to different atomic energy levels, characterizing bit-error and symbol-error rates to guide future interference management strategies. In FHSS, Sun \textit{et al.} \cite{sun2021interference} designed a receiver that uses interference for spectral shaping. By employing high-speed switchable delay lines and electro-optical sampling, their architecture achieves high shaping resolution and selectivity for fast-hopping anti-jamming communications.

\subsection{High-Frequency and Low-Frequency Communications}

%RAQ receivers cover a broad operating frequency range from DC to THz. At low frequencies, they enable detection beyond the wavelength-dependent size constraints of conventional antennas. At high frequencies, the pursuit of large bandwidth and high spectral efficiency has directed research interest toward multi-carrier techniques. Meanwhile, the limitation in instantaneous bandwidth motivates the exploration of FHSS.

RAQ receivers provide broad frequency accessibility for communication techniques across otherwise challenging spectra. At low frequencies, they enable detection beyond the wavelength-dependent size constraints of conventional antennas. At high frequencies, the pursuit of large bandwidth and high spectral efficiency has directed research interest toward multi-carrier techniques. Meanwhile, the wide tunable range and transition-selective response of RAQ receivers make FHSS attractive.

For low-frequency communications, the major mechanism lies on the AC stark shifts, as shown in Fig.~\ref{fig: advanced technologies}c.  Wang \textit{et al.} \cite{wang2025measurement} measured 250 kHz arc signals through the energy-level shifts from stark effect. Arumugam \cite{arumugam2025stark} extended this approach to the sub-kHz regime, achieving a sensitivity of $7.8\ \mu\text{V}/ {cm}^{-1}/\sqrt{\text{Hz}}$ at 300 Hz with Stark-modulated Rydberg states. Concurrently, Xie \textit{et al.} \cite{xie2025low} validated low-frequency communication by demodulating BPSK, OOK, and 2FSK signals near 100 kHz with an integrated vapor cell setup. For specialized applications, Xiao \textit{et al.} \cite{xiao2025optical} explored ULF field measurements tailored for power system diagnostics. A key technical challenge in these low-frequency experiments is the screening effect within the vapor cell, which can be addressed through cell designs and bias-field engineering. Jau \textit{et al.} \cite{jau2020vapor} utilized optically induced bias field in a sapphire cell, achieving a low-frequency noise floor of $0.34\ \text{mV}/ {cm}^{-1}/\sqrt{\text{Hz}}$, while Lei \textit{et al.} \cite{lei2024high} optimized DC bias fields to attain sensitivities of $5.7$, $2.2$, and $0.95\ \mu\text{V}/{cm}^{-1}/\sqrt{\text{Hz}}$ at 1, 10, and 100 kHz, respectively, outperforming small traditional dipole antennas.

The detection of high-frequency signals typically relies on the EIT-AT effect. To achieve large bandwidth and high spectral efficiency, current research focuses on multi-carrier techniques, as shown in Fig.~\ref{fig: advanced technologies}d. Du \textit{et al.} \cite{du2022realization} demonstrated dual-channel reception at 12.52 GHz and 39.80 GHz using different Rydberg states, achieving approximately 50 dB dynamic range and over 10 MHz bandwidth. Cui \textit{et al.} \cite{cui2025rydberg} advanced the theoretical foundations of multi-band RAQ receiver by deriving the closed-form transfer functions and revealing their roles as both atomic mixers and amplifiers. Kim \textit{et al.} \cite{kim2025multi} proposed a multi-band quantum wireless sensing method that employs a biased Gerchberg-Saxton algorithm to recover phase information, which enables AoA detection for multi-band users through correlation with multi-band steering vectors. For ultra-wideband multi-channel reception, Prajapati \textit{et al.} \cite{prajapati2024multichannel} employed optical frequency comb technology to enable massively parallel data acquisition in the AT regime without laser scanning, achieving a sensitivity of $2.3\ \mu\text{V}/{cm}^{-1}/\sqrt{\text{Hz}}$. Liu \textit{et al.} \cite{liu2022deep} developed a deep learning-enhanced RAQ receiver for multi-frequency microwave recognition, directly decoding FDM signals while reducing noise impact without explicitly solving the complex Lindblad master equation.

The broad frequency accessibility and transition-selective response of RAQ receivers make FHSS a natural approach for spectrum-agile communication, as shown in Fig.~\ref{fig: advanced technologies}e \cite{Qiu2026swfh}. Wen \textit{et al.} \cite{wen2024rydberg} experimentally demonstrated a verified RAQ-based frequency-hopping communication receiver based on a five-level atomic system and coherent population trapping effects, which realizes 62 dB dynamic range, 1 Mbps data rate, 50 MHz tunable bandwidth, 700 kHz instantaneous bandwidth, and a maximum hopping rate of 20,000 hops per second. In THz band, Nallappan and Skorobogatiy \cite{nallappan2022photonics} developed a photonics-based FHSS system operating in the 110-170 GHz window, transmitting 6 Gbps NRZ signals with protection against single- and multi-tone jamming attacks. Rostampoor and Adve \cite{rostampoor2025interference} demonstrated ultra-wideband frequency hopping covering 100 kHz to 20 GHz with 10 khop/s hopping rates and over 150,000 channels. To strengthen anti-jamming and anti-interception, Sun \textit{et al.} \cite{sun2021interference} proposed  an FHSS receiving scheme based on interference and frequency-to-time mapping, employing high-speed switchable delay lines and electro-optical sampling to generate eight receiving passbands within 25-42 GHz with significantly improved receiving selectivity.

\subsection{Multiple-Input Multiple-Output Communication}

\begin{figure*}[!t]
	\centering
	\includegraphics[width=\textwidth]{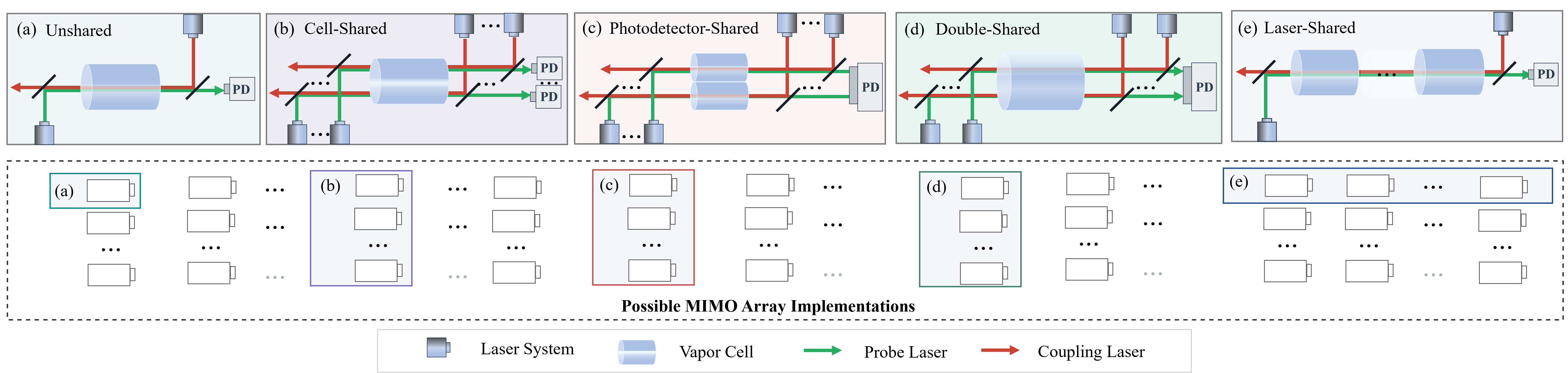}
	\caption{MIMO array based on RAQ receivers and different kinds of array element. (a) Conventional arrays employ no component sharing. (b) Photodetector-shared architecture. (c) Cell-shared architecture. (d) Double-shared architecture. (e) Laser-shared architecture.}
	\label{fig_MIMO}
\end{figure*}

Configuring RAQ receivers in a MIMO architecture can expand the effective aperture and spatial degrees of freedom of the receiving system. Recent studies mainly follow two directions, designing hardware-efficient atomic vapor-cell arrays and developing advanced signal processing methods to fully exploit the potential of atomic receiver arrays \cite{atapattu2026multishot,jeon2026fris}.

Expanding the effective aperture requires arraying multiple receiving units, each typically consisting of a vapor cell, a laser system, and a photodetector. Since fully independent replication introduces significant hardware complexity, RAQ MIMO research has concentrated on component-multiplexing topologies. Wu \textit{et al.} \cite{wu2025low} provided a foundational taxonomy based on vapor-cell and photodetector sharing. As shown in Fig.~\ref{fig_MIMO}a, in the conventional unshared array, each element acts as an independent receiver.
%{\color{red}Studies using this configuration have primarily advanced the signal processing theory for atomic MIMO, including formulating signal detection as a nonlinear biased phase retrieval problem \cite{cui2025towards, cui2024multi}, framing it within a hypothesis testing framework \cite{atapattu2025detection}, deriving fundamental capacity bounds \cite{gong2025rydberg, yuan2025electromagnetic}, and developing CSI-free detection methods \cite{song2025csi}. Precoding techniques to approach these capacity limits have also been a key focus \cite{cui2025mimo, zhu2025raq}.} 
To reduce system size, the cell-shared architecture in Fig.~\ref{fig_MIMO}b allows multiple laser beams to probe a common vapor cell, and and has been applied to multi-user MIMO uplink with reduced power consumption \cite{gong2025rydberg}. The photodetector-shared architecture in Fig.~\ref{fig_MIMO}c combines optical outputs from multiple cells onto a single detector, and relies on digital signal processing for channel separation. The double-shared configuration in Fig.~\ref{fig_MIMO}d offers more compact footprint by sharing both cell and detector, but increases demultiplexing complexity. In addition, several studies proposed laser-shared configurations that enable multiple receiving channels within a single RAQ receiver \cite{cui2026singleantenna,cui2026continuous,guo2025aoa}, as shown in Fig.~\ref{fig_MIMO}e. By partitioning the atomic medium into spatially separated segments, it extends the effective RF-atom interaction length without increasing the total optical propagation distance, thereby preserving transmission fidelity. Notably, this laser-shared approach can be integrated with the above configurations to form hybrid multiplexing schemes with higher component reuse and hardware efficiency.

Advanced signal processing and system optimization techniques further enable these array architectures to address the inherent nonlinearity and phase-loss issues of RAQ receivers. For signal detection, Cui \textit{et al.} formulated MIMO detection as a nonlinear biased phase retrieval problem and proposed enhanced Gerchberg-Saxton algorithms \cite{cui2025towards, cui2024multi}. Similarly, Atapattu \textit{et al.} \cite{atapattu2025detection} designed robust detectors within a hypothesis testing framework, while Liu \textit{et al.} \cite{Liu2026prss} introduced phase-rotated symbol spreading (PRSS) to reconstruct an effective linear model without spectral efficiency loss. To bypass explicit channel state information (CSI) estimation, Song \textit{et al.} \cite{song2025csi} employed in-context learning for direct symbol recovery. Regarding system optimization, precoding techniques such as IQ-aware digital precoding \cite{cui2025mimo} and quantum-weighted MMSE \cite{zhu2025raq} have been proposed to maximize multi-band spectral efficiency and mitigate interference. Kang \textit{et al.} \cite{Kang2025deepq} developed DeepQ-MIMO to solve nonconvex joint transceiver optimization through deep learning with reference signal injection. Gong \textit{et al.} \cite{gong2025rydberg} and Yuan \textit{et al.} \cite{yuan2025electromagnetic} confirmed that these algorithmic advancements enable RAQ MIMO systems to approach capacity bounds, while achieving power savings and rate gains over conventional antennas.

\section{Application Scenarios, Open Challenges, and Future Prospects}

\subsection{Typical Communication Application Scenarios}
\subsubsection{Satellite and Space Communications}

RAQ receivers have emerged as a promising technology for satellite communications and navigation by overcoming the limitations of conventional RF front ends, including propagation loss and stringent SWaP constraints. Peng \textit{et al.} \cite{peng2025rydberg} proposed an RAQ-MIMO satellite architecture and demonstrate a squaring gain over conventional RF MIMO under Rayleigh fading, leading to smaller antenna apertures, lower transmit power, and extended communication ranges. Lei \textit{et al.} \cite{lei2025satellite} reported the first direct reception of geostationary satellite signals using an enhanced Rydberg receiver without a low-noise amplifier. Moreover, hybrid atomic-electronic satellite receivers have been predicted to provide spectral efficiencies exceeding 6 bit/s/Hz and coverage extensions of up to 1000 km \cite{peng2025rydberg}. In addition, the quantum spatially encoded non-stationary effect has been proposed to enable instantaneous two-dimensional localization using a single satellite equipped with a single RAQ receiver, outperforming conventional antennas in both accuracy and the required number of satellites \cite{guo2026instantaneous}.

\begin{figure*}[!t]
	\centering
	\includegraphics[width=1\textwidth]{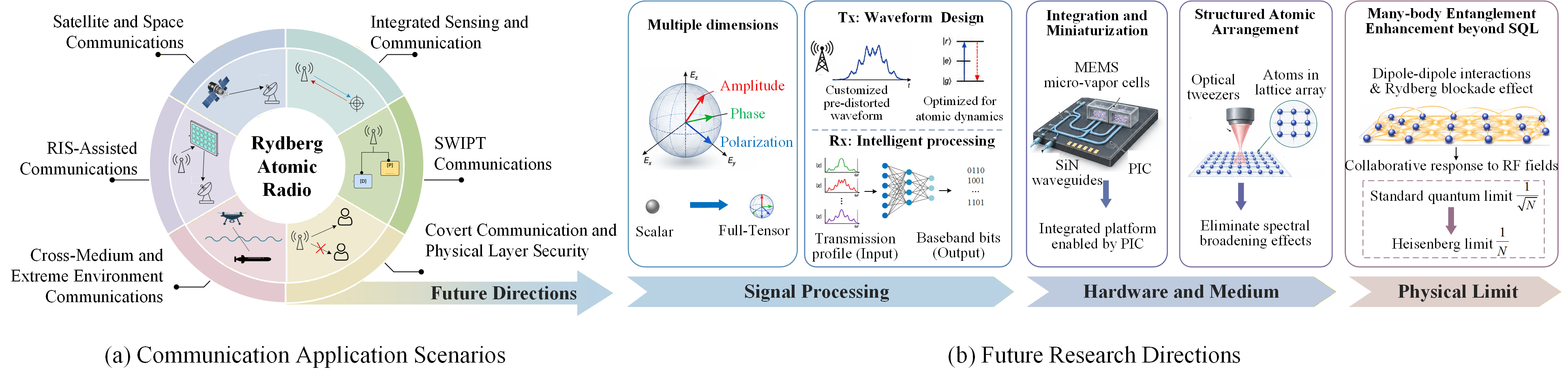}
	\caption{Typical communication application scenarios and future research directions.}
	\label{fig_applications&direcetions}
\end{figure*}

\subsubsection{Integrated Sensing and Communication}

The ultra-wide band tunability and high sensitivity of RAQ receivers make them promising platforms for integrated sensing and communication. Arumugam \textit{et al.} exploited XM satellite signals of opportunity for soil-moisture sensing, demonstrating the potential of RAQ receivers for passive remote sensing \cite{arumugam2024remote}. In communication, Yuan \textit{et al.} experimentally demonstrated a 24.12 GHz Rydberg receiver for 5G millimeter-wave communications, achieving an 82.5 kbit/s data rate \cite{yuan2023rydberg}. Bussey \textit{et al.} further showed that Rydberg receivers can simultaneously recover communication signals and track RF signal fading and frequency drift, enabling joint communication and channel-state sensing \cite{bussey2021rydberg}. More recently, Jeon and Chae developed a reconfigurable intelligent surface (RIS)-aided ISAC framework with multiple Rydberg receiver users, establishing a unified communication-sensing model and a joint optimization framework for future 6G systems \cite{jeon2026ris}.

\subsubsection{Reconfigurable Intelligent Surface-Assisted Networks}

RAQ receivers are increasingly being integrated with reconfigurable intelligent surfaces to create high-performance and adaptive wireless links. Peng \textit{et al.} \cite{peng2025ris} investigated a RIS-assisted atomic MIMO receiver architecture, where pulse amplitude modulation is employed to effectively mitigate phase ambiguity by aligning the transmitted signal phase with the LO, improving signal detection performanc. Jeon and Chae  \cite{jeon2026ris} further developed a RIS-aided ISAC framework with multiple RAQ receiver users, jointly optimizing communication performance and sensing accuracy under CRB constraints.

\subsubsection{Simultaneous Wireless Information and Power Transfer}

The high sensitivity of RAQ receivers opens new opportunities for simultaneous wireless information and power transfer (SWIPT) networks, particularly for supporting battery-free IoT devices. Peng \textit{et al.} \cite{peng2025from} proposed a hybrid SWIPT-MIMO architecture that enables reliable uplink communication from energy-harvesting IoT devices using a RAQ receiver, providing a viable pathway toward battery-free wireless networks.

\subsubsection{Cross-Medium and Extreme Environment Communications}

The ultra-wide frequency tunability of RAQ receivers makes them attractive for cross-medium communications across air, terrestrial, maritime, and underwater environments. Unlike conventional receivers that require different apertures and front ends for different frequency bands, RAQ receivers can support a wide spectral range using the same atomic front end. This is particularly promising for future SAGSIN, where seamless communication across heterogeneous media and extreme environments is required.

\subsubsection{Covert Communication and Physical Layer Security}

The high sensitivity, frequency agility, and atom-resonance selectivity of RAQ receivers also provide new opportunities for covert communication and physical-layer security. Their ability to detect weak signals can reduce the required transmit power, while their tunable and selective atomic response can support low-probability-of-intercept reception and interference-resilient operation. Yu \textit{et al.} \cite{yu2025covert} proposed a RAQ-enabled covert semantic communication scheme, demonstrating improved covertness while preserving transmission reliability. 

\subsection{Open Challenges and Future Research Directions}
Despite the immense potential and recent progress of RAQ receivers, this technology is still in its nascent stage, with most studies focusing on feasibility verification of field sensing and signal reception. To transform RAQ receivers from laboratory to practical communication systems, there remains several open challenges that need to be considered.
%Despite the significant potential and recent progress of RAQ receivers, this technology is still in its early stage, with most existing studies focusing on feasibility demonstrations of field sensing and signal reception. To advance RAQ receivers from laboratory prototypes toward practical communication systems, several open challenges must be carefully addressed.

\begin{itemize}
	\item \textit{Practical Performance Gap}: Although RAQ receivers show theoretical advantages in sensitivity, detectability, and frequency coverage, current experimental systems remain far from these limits. Practical performance is constrained by laser stability, intensity noise, vapor-cell temperature, atomic density, thermal motion, Doppler broadening, detector noise, and other environmental perturbations. These factors degrade sensitivity, bandwidth, phase stability, and long-term robustness. %Moreover, most existing systems still rely on scalar field-strength detection, so further gains may require exploiting additional dimensions such as phase, polarization, spatial distribution, and structured-field modes.
	
	\item \textit{Bandwidth Bottleneck}: RAQ receivers can in principle access frequencies from DC to THz, but practical operation is constrained by discrete energy-level spacings, accessible transitions, laser tuning ranges, and state-preparation conditions. Thus, wide spectral accessibility does not imply continuous and uniformly sensitive frequency coverage. Meanwhile, the finite EIT or AT response time limits the instantaneous bandwidth, directly restricting high-speed reception, broadband waveform transmission, and achievable communication capacity.
	
	\item \textit{Nonlinear and Environment-Sensitive Modeling}: Modeling RAQ receivers is challenging as their input-output relationship is derived from nonlinear atom-field dynamics governed by the Hamiltonian and master equations, rather than from a simple linear channel assumption. Even in ideal conditions, finite dynamic range and saturation effects may cause gain compression and intermodulation distortion under strong or multi-tone excitation. In practical systems, Doppler averaging, detuning, device imperfections, and environmental drift further perturb the atomic response. These coupled effects make theoretical models difficult to maintain accurately, posing challenges on waveform design, demodulation, and performance analysis.
	﻿

	\item \textit{Deployment Barrier of Optical Hardware}: Current RAQ receivers still rely on laboratory-scale optical hardware, including discrete lasers, frequency-stabilization modules, vapor cells, optical paths, photodetectors, temperature-control units, and readout electronics. Although atomic receivers can overcome the wavelength-scale aperture constraint of conventional antennas, the complete system remains bulky, power-consuming, costly, and sensitive to vibration, temperature drift, and optical misalignment. %Practical deployment requires compact lasers, integrated photonics, chip-scale vapor cells, low-power stabilization, robust packaging, and scalable atomic-array architectures.
	
\end{itemize}

Addressing these challenges requires
%Bridging the gap between laboratory systems and practical high-performance communication platforms requires 
coordinated advances in the in-depth exploration of physical mechanisms and the maturation of engineering integration across multiple dimensions. In this context, six key future research directions for RAQ receivers are provided as follows. %including multi-dimensional signal reception, tailored waveform design, intelligent receiver processing, structured atomic arrays, miniaturized hardware integration, and pure quantum-limited detection.

\subsubsection{Full-Tensor Polarization Sensing and Multi-Dimensional Multiplexing}
Current RAQ receivers primarily extract the scalar amplitude of RF electric fields from EIT or AT splitting. The physical properties related to the real part, i.e., dispersion response, and the spatial polarization resolution brought by tensor polarizability remain underexplored. Future research should exploit the full-tensor polarizability and coherent optical response of Rydberg atoms to simultaneously measure the amplitude, phase, and orthogonal polarization states of RF fields. This capability enables vector electromagnetic field sensing and facilitates polarization- and spatial-domain multiplexing, thereby improving spectral efficiency and channel capacity.

\subsubsection{Atom Dynamics-Tailored Waveform and Modulation Schemes Design}
Existing studies primarily focus on receiver design, which directly adopt communication waveforms and modulation schemes developed for conventional electronic RF receivers, without accounting for the unique characteristics of atoms. 
%Inevitably, the continuous evolution of RAQ receivers calls for transmitter-receiver co-design tailored to the atomic dynamic response. 
Future research can focus on the nonlinear predistortion for waveform optimization to mitigate atom-induced signal distortions, as well as novel orthogonal signal bases with dedicated modulation schemes to match the atomic evolution trajectories. Such designs may improve spectral efficiency, alleviate instantaneous bandwidth limitations, and unlock the full communication potential of RAQ receivers.

\subsubsection{AI-Empowered RAQ Processing and End-to-End Demodulation}
%In practical environments, the interactions among atoms, laser fields, and RF fields exhibit highly nonlinear and time-varying characteristics. Demodulation schemes based on semi-classical Hamiltonian models are prone to model mismatch, insufficient accuracy, or even complete failure.

To bypass the complicated physical modeling and overcome model mismatch issue under various environmental interference, 
future work can integrated lightweight artificial intelligence (AI) algorithms to implement nonlinear equalization for dynamic range expansion, quantum-aware denoising and adaptive optimal operating point tracking. More importantly, end-to-end learning frameworks that directly map atomic spectra to baseband digital signals can be explored to simplify the processing pipline. Further, the integration with semantic communications that recovery signal directly from the atomic response may enable more reliable  communication in strong-interference and low-SNR scenarios.

\subsubsection{Photonic Integrated Circuits and Microcavity-Enhanced Miniaturization}
%To date, experimental validations of RAQ receivers mostly rely on free-space optical setups built with discrete components and frequency-stabilized laser, resulting in large size and limited environmental adaptability.
To address the deployment barrier of laboratory optical hardware, 
future receiver designs are expected to evolve toward monolithic photonic integrated circuit (PIC)-based architectures that integrate optical functionalities and atomic vapor cells into compact platforms. Furthermore, optical microcavities can be employed to strengthen atom–light interactions and maintain a high-SNR quantum coherent state even within an extremely small effective volume, ultimately paving the way for chip-scale RAQ receivers front-end.

\subsubsection{Optical Tweezers for Structured Atomic Arrangement and Atom Array}
%Currently, RAQ receivers mainly employ thermal atomic ensembles, where random atomic distribution and thermal motion cause Doppler and transit-time broadening, restricting instantaneous bandwidth and sensitivity of the system. 
To suppress thermal atomic spectral broadening, 
a promising direction is spatially ordered atomic arrays with precisely controlled inter-atomic spacing, which can be achieved through laser cooling and optical tweezer technologies. The inter-atomic spacing can be controlled at the micrometer scale with high positional accuracy, which is still far below the half wavelength of RF signals. Such structured atomic arrangements could suppress spectral broadening effects, overcome the grating lobe limitations in traditional phased arrays, and enable quantum antennas for distortion-free beamforming, ultra-high-precision phase and DoA estimation.

\subsubsection{Atomic Many-Body Entanglement Enhanced Reception Beyond the SQL}
Under the semi-classical physical picture, the sensitivity is ultimately bounded by the standard quantum limit (SQL), scaling as $\propto 1/\sqrt{N}$. One potential approach to surpassing this limit is exploiting dipole–dipole interactions and the Rydberg blockade effect to prepare highly entangled many-body quantum states that respond collectively to RF fields. Such entangled ensembles could approach the Heisenberg limit, i.e., $\propto 1/N$, enabling communication reception beyond classical constraints and supporting quantum-limited communications under ultra-long-distance and ultra-weak-signal.

\section{Conclusion}
RAQ radio that shifts the wireless reception paradigm from classical field-to-current conversion to atomic quantum state mapping, offers a promising path to transcend longstanding electronic bottlenecks. 
From a wireless communication perspective, this survey have systematically synthesized the RAQ radio along two dimensions. Mechanistically, we reviewed the representative system architectures, physical response models, and equivalent channel formulations that define the capacity boundaries of RF-to-optical transduction. Application-wise, we have analyzed state-of-the-art performance enhancement techniques alongside inherent trade-offs of key metrics, and detailed RAQ-enabled advanced communication technologies, including intelligent cognitive communication, anti-interference communication and full-spectrum reception.  We also discussed emerging application scenarios such as satellite communication and integrated sensing and communications.
By clarifying the physical mechanisms, communication models, enabling techniques, and open challenges of RAQ radio, this survey provides a useful reference for future research on deployable atomic receivers and quantum-enhanced wireless communications.

\bibliography{IEEEabrv,references}

\end{document}